\documentclass[11pt]{JHEP}
\usepackage{amsmath}
\usepackage{amssymb}
\usepackage{amsmath,amsthm,epsfig,euscript,array,cancel}
\font\mybb=msbm10 at 10pt
\def\bb#1{\hbox{\mybb#1}}

\newcommand{\be}{\begin{equation}}
\newcommand{\ee}{\end{equation}}

\def\bea{\begin{eqnarray}}
\def\eea{\end{eqnarray}}
\preprint{June 19, 2014. V2: August 6, 2014}
\renewcommand{\theequation}{\arabic{section}.\arabic{equation}}
\title{On Lagrangian  approach to self-dual gauge fields in spacetime of nontrivial topology}

\author{Igor Bandos
 ~ \\ {\small\it UPV/EHU, P.O. Box 644, 48080 Bilbao, Spain}  ~\\ and ~ \\ {\small\it
IKERBASQUE, Basque Foundation for Science, 48011, Bilbao, Spain} }

\date{ 14/06/2013--19/06/2014, V2--16/07/2014, V2'--06/08/2014,  Printed \today }

\abstract{We study the Lagrangian description of chiral bosons, p-form gauge fields with
(anti--)self-dual gauge field strengths, in  $D=2p+2$ dimensional spacetime of nontrivial topology. We show that the manifestly Lorentz and diffeomorphism invariant
Pasti--Sorokin--Tonin (PST) approach is consistent and produces the (anti-)self-duality equation also in  topologically nontrivial spacetime. We discuss in what circumstances the nontrivial topology makes difference between two
disconnected, {\it da-timelike} and {\it da-spacelike} branches of the PST system, the gauge fixed version of which are described by not manifestly invariant Henneaux-Teitelboim (HT) and Perry--Schwarz (PS) actions, respectively.}


\begin{document}

\section{Introduction}

The PST (Pasti--Sorokin--Tonin) approach \cite{Pasti:1995tn,Pasti:1996vs} provides a manifestly Lorentz invariant Lagrangian description of the self-dual gauge fields as well as of more general duality-invariant theories. One of the simplest examples is provided by D=6  theory of chiral 2-form potential $B_2$, where the PST action allows to reproduce the self-duality or anti-self duality condition $H_3=\pm *H_3$ for the field strength $H_3=dB_2$. It is also  known the Dirac-Born--Infeld type version of this model producing the nonlinear generalization of the self-duality conditions and describing the M-theory 5-brane (M5-brane) in its bosonic limit \cite{Pasti:1997gx} and in a complete form \cite{Bandos:1997ui,Aganagic:1997zq}.

The PST action involves an auxiliary scalar field $a(x)$ which enters the Lagrangian only through its derivatives
$\partial_\mu a(x)$. Hence its shift by constant parameter is a (Pecci-Queen) symmetry. But moreover, the PST
scalar $a(x)$ is a St\"uckelberg field as far as the PST action possesses a gauge symmetry (so-called {\it
second PST} or {\it PST2} symmetry) which can be `parametrized' by (almost) arbitrary variations $\delta a(x)$
of $a(x)$.  The square of derivatives of this PST scalar enters the denominator of the Lagrangian so that it is
not allowed to be zero or constant, but it can be gauged to coincide with {\it e.g.} one of the space or time
coordinates (but not with their light-like combinations $x^{\pm}=t\pm x$). Fixing $da=dt$ in the PST Lagrangian
for 6D chiral bosons we  arrive at the non-manifestly Lorentz invariant Henneaux-Teitelboim (HT) action
\cite{Henneaux:1988gg}; while fixing, say, $da=dx^1$, we arrive at the Perry--Schwarz--type  (PS) action
\cite{Perry:1996mk}.

As we will discuss in the main text,  the gauges $da=dt$ and, say, $da=dx^1$ cannot be connected by a
nonsingular PST2 gauge transformations. Thus the PST action actually describes a dynamical system with two
branches, where, respectively, the gauge   $da=dt$ and the gauge $da=dx^1$ can be fixed; let us call these {\it da--timelike} and {\it da-spacelike} branches of the {\it PST-system}.  In topologically trivial situation these branches  are classically equivalent in the sense that both can be used to produce the (anti--)self-duality as a gauge fixed version of the equations of motion.

 In this paper we study the PST system in spacetime of nontrivial
topology. The standard PST action is well defined if the spacetime admits a nowhere vanishing vector field.
This is always the case in the spacetime with metric of Lorentzian signature.
To be more explicit, generically the nowhere vanishing {\it timelike} vector field exists in it;
this can be related  with the exterior derivative of the PST scalar $da$ thus allowing for a {\it
$da$--timelike} branch of the PST system (and for its gauge fixed version, the HT action), while to have also
{\it $da$--spacelike} branch (and its gauge fixed version, the PS action), we need the Lorentzian
spacetime to admit the second, spacelike  nowhere vanishing vector field.

In a topologically nontrivial  spacetime with Euclidean metric which does not admit nowhere vanishing vector
field, the standard PST construction is not well defined. However, if it allows for existence of a nowhere vanishing 'q-plane field' with $q>1$, then a modified version of the PST action with several ($q$) PST scalars
\cite{Pasti:2009xc,Ko:2013dka} can be constructed. These cases (which are associated with different
$D=q+(D-q)$ splittings of spacetime \cite{Belov:2006jd,Ho:2008nn,Chen:2010jgb}) are beyond the scope
of this paper.

The standard PST action possesses one more gauge symmetry, the so-called {\it PST1 symmetry}, which leaves
inert the PST scalar $a(x)$ but plays a very important role in derivation of the self-duality (or duality)
equation. The point is that the variation of the PST action produces (a special type of the) second order
equation, and the self-duality equation can be deduced from it by using the gauge fixing of the PST1 symmetry.
In the original discussion of \cite{Pasti:1995tn,Pasti:1996vs}, as well as in all presently known applications of PST
technique (see e.g. \cite{Dall'Agata:1997db,Bandos:1997gd,Dall'Agata:1997ju,Dall'Agata:1998va,BSS2013}),  this
procedure included the stage of  solving the condition for some p-form to be closed, $dG_p=0$, by writing it as
an exact form, $G_p=dP_{p-1}$. Then one can show that $P_{p-1}$ can be gauged away by the PST1 symmetry and
that the remaining equation $G_p=0$ is equivalent to the self-duality equation for the field strength of the
gauge potential.

The procedure described above is literally applicable to the case of flat  or topologically trivial spacetime.
To be more precise, it is valid when the $p$-th Betti number of the spacetime manifold $M$ vanishes, $b_p=dim
({\bb H}^{p}(M, {\bb R})=0$. However, if $b_p\not=0$, there exists a (set of) closed but not exact $p$--form(s)
$\omega_p^\Lambda$, $\Lambda=1,...,b_p$ so that the general solution of  $dG_p=0$ reads
$G_p=\sum\limits_{1}^{b_p}c_\Lambda\omega_p^\Lambda + dP_{p-1}$ with constant $c_\Lambda$. (This is the place
to stress that our discussion here is schematic, and actually in some cases we will arrive at the expressions with
functions $l_\Lambda(a(x))$ of the PST scalar $a(x)$ instead of constants $c_\Lambda$). At the first glance,
this seems to spoil the derivation of the self-duality equation from the PST Lagrangian.

The main aim of the present paper is to show that this is not the case: the PST approach  is pretty consistent
and is able to produce the wanted (anti--)self-duality equations also in the case of topologically nontrivial
spacetime.

To see this one has to notice the presence of an unusual type of symmetry parametrized by function(s) of the
PST scalar, $f(a(t,\vec{x}))$,  which we call ''semi-local symmetry''\footnote{When this paper was finished the author have become  aware that for the noncovariant HT action in ${\bb R}\otimes M^5$ spacetime with $b_2(M^5)\not=0$ such a symmetry had been found in \cite{Bekaert:1998yp}. The author thanks Marc Henneaux for having brought this paper to his attention.}. In the {\it da-timelike} branch, where
the PST scalar can be gauge fixed to coincide with the time coordinate,  this semi-local symmetry is a gauge
symmetry which  can be  used in the above mentioned derivation of the self-duality equation. In contrast, in the {\it da-spacelike}  branch, where $a(x)$ can be gauge fixed to
coincide with one of the  special coordinate,  the semi-local symmetry is an infinite dimensional local
symmetry, similar to the two dimensional conformal symmetry. This cannot be used to fix any gauge and, as a
result, in some topologically nontrivial situations the self--duality equation cannot be produced in this
branch.

Thus, interestingly enough, the nontrivial spacetime topology singles out the {\it da-timelike} branch of  the
PST system.

The rest of this paper is organized as follows. In Sec. 2 we review the PST approach to Lagrangian description
of the chiral bosons in flat six dimensional spacetime. We obtain Lagrangian equations of motion form the PST
action for two-form gauge potential $B_2$, describe the gauge symmetries of this action, the set of which
includes the so-called PST1 and PST2 gauge symmetries, and show that these allow to fix a gauge where the
Lagrangian equations of motion reduce to the anti-self-duality equation.

In Sec. 3 we elaborate this formalism in topologically nontrivial spacetime $M^{6}=M^{1+5}$ and prove its
consistency. In sec. 3.1. we obtain the first order form of the PST Lagrangian equations and find that in
certainly topologically nontrivial spacetimes, including $M^{1+5}$ with  $b_2\not=0$, this contain additional
topological contribution defined by a number of functions of one variables. In sec. 3.2. we show that in such
spacetimes, in addition to the PST1 and PST2 gauge symmetries, the PST action also possesses an additional
semi-local symmetry and that the additional topological contributions to the first order form of the
Lagrangian equations can be compensated or generated by transformations of this semi-local  symmetry. In sec.
3.3. we show that in the $da$-timelike branch of the PST system, as well as in its gauge fixed version described
by HT action, the semi-local symmetry is a gauge symmetry, while in the da-spacelike branch of the PST system
and in its gauge fixed version described by the PS action, this is an infinite dimensional rigid symmetry (similar to 2d conformal symmetry).
 This allows us to derive (in sec. 3.4.) the anti--self--duality equations as gauge fixed version of the
 Lagrangian equations of motion which follows from the $da$-timelike PST action and HT action.

In Sec. 4 we discuss the two dimensional PST action for the chiral bosons, where the counterpart of semi-local
symmetry is present also in the case of topologically trivial 2d spacetime. The case of chiral bosons in a
topologically nontrivial spacetime of arbitrary even dimension $D=2p+2$ is addressed  in Sec. 5. A strong
similarity with 6D case allows us to be short and to restrict ourselves by presenting the basic equations
 and formulating the results for this general case.  We conclude in Sec. 6.

In appendix A we give additional technical details on derivation of the first order form of the PST Lagrangian
equations in flat spacetime. In Appendix B we discuss the properties of Noether currents for gauge symmetries
on a specific  example of 2-form gauge fields theories in D=6.

For the reader convenience we have make the presentation of D=6, of D=2 and of the general D=2p+2 cases self-sufficient, so that each of these three can be read separately; the price to pay was repetition of the basic statements, although with some specific modifications.

\bigskip

{\it Abbreviations:}
\begin{itemize}
\item PST is used for Pasti-Sorokin-Tonin approach \cite{Pasti:1995tn,Pasti:1996vs,Pasti:1997gx}. The PST action for
    2-form gauge field in D=6 can be found in Eq. (\ref{LHH=}); see (\ref{LPST=D}) for the case of $p$-form
    gauge fields in $M^{2p+2}$. The PST action for 2d chiral bosons \cite{Maznytsia:1998xw} is presented in
    (\ref{cL}). \item HT is used for Henneaux-Teitelboim action (\ref{L6HT:=}) \cite{Henneaux:1988gg}.
    \item PS is used for Perry--Schwarz action (\ref{L6PS:=}) \cite{Perry:1996mk}. \item  FJ is used for
    Floreanini--Jackiw action \cite{Floreanini:1987as} which can be found in Eq. (\ref{SFJ=}).
\end{itemize}

\section{PST Lagrangian for 6D selfdual gauge fields}

\subsection{ 2-form gauge field in D=6}

\subsubsection{Free action for  2-form gauge potential in $M^{1+5}$}

We begin by discussing the theory of free non-chiral 2-form gauge potential
\begin{eqnarray}\label{B2:=}
B_2={1\over 2}dx^\nu\wedge dx^\mu B_{\mu\nu}(x)
\end{eqnarray}
in six dimensional spacetime.   In Eq. (\ref{B2:=}) $x^\mu$ are local coordinates of $M^6$,  $\mu,  \nu
=0,1,..., 5$, and  $\wedge$ denotes the exterior product of differential forms, $dx^\mu\wedge dx^\nu=-
dx^\nu\wedge dx^\mu$, so that the second rank tensor $B_{\mu\nu}(x)$ contributing to (\ref{B2:=}) is
antisymmetric,  $B_{\mu\nu}(x)=- B_{\nu\mu}(x)$. The standard field strength of the 2-form gauge potential is
\begin{eqnarray}\label{H=dB}
H_3={1\over 3!} dx^\rho \wedge dx^\nu\wedge dx^\mu  H_{\mu\nu\rho}(x):= dB_2
\end{eqnarray}
where $d=dx^\mu\partial_\mu$ is the exterior derivative which in our notation acts from the right, {\it e.g.}
\begin{eqnarray}\label{dB2:=}  dB_2 &=& {1\over 2}dx^\mu\wedge dx^\nu \wedge dx^\rho \partial_\rho
B_{\nu\mu}(x) \nonumber \\ &&  \equiv {1\over 3!}dx^\mu\wedge dx^\nu \wedge dx^\rho \left(\partial_\rho
B_{\nu\mu}(x)+
\partial_\mu B_{\rho\nu}(x)+ \partial_\nu B_{\mu\rho}(x)\right)\; , \qquad
\end{eqnarray}
so that (\ref{H=dB}) implies  $H_{\mu\nu\rho}= 3 \partial_{[\mu} B_{\nu\rho ]}:=
\partial_{\mu} B_{\nu\rho } + \partial_{\nu} B_{\rho\mu}+ \partial_{\rho } B_{\mu\nu}$.

The free field action for the 2-form gauge potential reads
\begin{eqnarray}\label{S0=H*H}
S_0= \int_{M^6}{\cal L}^0\; , \qquad {\cal L}^0= {1\over 2} H_3\wedge *H_3\; , \qquad
\end{eqnarray}
where the Lagrangian 6-form  ${\cal L}^0$ is written using the Hodge star symbol,
\begin{eqnarray}\label{*H:=}
*H_3:= {1\over 3!}dx^\rho\wedge dx^\mu \wedge dx^\nu (*H)_{\mu\nu\rho}\; ,\qquad (*H)^{\mu\nu\rho}:= {1\over 3!
\sqrt{|g|}} \varepsilon^{\mu\nu\rho\mu^\prime\nu^\prime\rho^\prime} H_{\mu^\prime\nu^\prime\rho^\prime} \; ,
\qquad
\end{eqnarray}
$\varepsilon^{\mu\nu\rho\mu^\prime\nu^\prime\rho^\prime}
=\varepsilon^{[\mu\nu\rho\mu^\prime\nu^\prime\rho^\prime]}$ is the Levi-Civita antisymmetric tensor density
normalized so that $\varepsilon^{012345}=1$ and $g=det (g_{\mu\nu})$ is the determinant of the spacetime metric
which in our conventions is of mostly minus signature.

Varying the action (\ref{S0=H*H}) with respect to the gauge potential, one finds the standard free
'Maxwell--like' equations for the 3-form field strength
\begin{eqnarray}\label{d*H=0}
d*H_3=0\; . \qquad
\end{eqnarray}
By construction,  the field strength also obeys the Bianchi identities,
\begin{eqnarray}\label{dH=0}
dH_3=0
 \; . \qquad
\end{eqnarray}

In spacetime of nontrivial topology, (\ref{H=dB}) generically gives only a particular solution of (\ref{dH=0}).
The general solution
\begin{eqnarray}\label{H=dB+}
H_3= dB_2 + k_L\Omega_3^L\; , \qquad  d\Omega_3^L=0\; , \qquad \Omega_3^L\not=d\Xi_2^L\; , \qquad L=
1,...,b_3\; , \quad
\end{eqnarray}
contains a topological  contribution $k_L\Omega_3^L$ determining the cohomological class $[H_3]$ of $H_3$ as
$[H_3]=k_L [\Omega^L]$. Here $\Omega_3^L$ is the basis of the 3rd cohomology group $ {\bb H}^3(M^6)$ of the
spacetime $M^6$, $b_3= dim {H}^3(M^6)$, and  $k_L$ are constant coefficients (see e.g.  \cite{Nash:1983cq}).

Despite the presence of topological contributions, the variational problem is usually defined within  a fixed
cohomology class (see \cite{Hitchin:2000jd} and e.g.  \cite{Belov:2006jd}), this is to say
\begin{eqnarray}\label{vH=dvB-2}
\delta H_3= d\delta B_2 \quad
\end{eqnarray}
with an arbitrary  $\delta B_2$.

\subsubsection{(Anti-)self duality equation}

A particular solution of the second order equation (\ref{d*H=0}) is provided by 3-form field strength obeying
the anti-self-duality condition
\begin{eqnarray}\label{H=*H}
H_3+ *H_3=0\qquad \Leftrightarrow \qquad  H_{\mu\nu\rho}+ { \sqrt{|g|} \over 3!}
\varepsilon_{\mu\nu\rho\mu^\prime\nu^\prime\rho^\prime} H^{\mu^\prime\nu^\prime\rho^\prime}=0 \; . \qquad
\end{eqnarray}
Clearly, this first order equation is not satisfied by  the most general solution of (\ref{d*H=0}).
Furthermore, the action (\ref{S0=H*H}) vanishes on this solution. (The above statements are also valid for  the
self-duality condition $H_3-*H_3=0$).

It was natural to wander whether it is possible to  construct  the action which produces just the (anti-)self
duality equation, or dualities between the field strengths of different ranks, as equations of
motion. The action principle of such a type was of great interest for the development of supersymmetry and
string theory as far as many important supermultiplets, such as the ones of 10D type IIB supergravity and of
the M-theory 5-brane, included self-dual and/or anti-self-dual tensor  fields. Such duality invariant actions with broken Lorentz symmetry (better to say, which are not-manifestly Lorentz invariant \cite{Maznytsia:1998xw}) were proposed in  \cite{Zwanziger:1970hk} and in \cite{Henneaux:1988gg,Perry:1996mk}. The covariant action  principle was
developed in \cite{Pasti:1995tn,Pasti:1996vs}
\footnote{Notice also the existence of a covariant  action principle with infinitely many
auxiliary fields \cite{Berkovits:1996nq,Bengtsson:1996fm} which describes just the gauge fields obeying
(anti-)self duality equations if the additional restrictions to the configurations containing only finite
number of fields is imposed. A quite curious nonlocal  approach to chiral bosons was proposed in \cite{Giannakis:1997hv}; it implies 1+5 splitting of 6d coordinates and the canonical brackets for 6d 2-form potentials is the same as in HT approach,  but its relation with HT action \cite{Henneaux:1988gg}  is not clear for us.
There exists also a pragmatic approach  in the frame of which the
(anti-)self-duality or/and duality equations do not follow from the action but are imposed by hand afterwards.
In such a way the 'democratic' formulation of 10D type IIA and IIB supergravities   \cite{Bergshoeff:2001pv}
and the alternative 5-brane action of \cite{Cederwall:1997gg} were constructed. However, for instance,
switching on interactions in a covariant approach of this type does not look so straightforward as when having the action principle of canonical type, like  \cite{Bandos:1997ui,Aganagic:1997zq,Dall'Agata:1997db,Bandos:1997gd,Dall'Agata:1998va}.  }. The
properties of this PST action in spacetime of nontrivial topology is our main interest in this paper.

\subsection{Duality invariant action for the 2-form gauge field. PST, HT and PS actions; $da$--timelike and
$da$--spacelike branches of PST system }

\subsubsection{PST action}
The standard PST Lagrangian 6--form
\begin{eqnarray}\label{LHH=}
{\cal L}^{PST}_6 = - ( i_v* H_3+ i_vH_{3}) \wedge H_{3}\wedge v &=& \nonumber \\ &=& {1\over 2} d^6 x \, \sqrt{
|g| }
 v^\rho\left(*H_{\mu\nu\rho} +     H _{\mu\nu\rho} \right) *H^{\mu\nu\lambda}\,v_\lambda \; , \qquad
\end{eqnarray}
is defined on any 6-dimensional manifold $M^6$ which allows for the existence of nowhere vanishing vector
field; it is real when spacetime has a metric of Lorentz signature,  $M^6=M^{1+5}$.

In (\ref{LHH=}) $x^\mu$ are local coordinates on $M^6$, $|g|=|det g_{\mu\nu}|$,
\begin{eqnarray}\label{v=da}
v=dx^\mu v_\mu = {da\over \sqrt{ |\partial a\partial a| }} \; , \qquad \partial a\partial a := \partial_\mu a\,
g^{\mu\nu}(x)\, \partial_\nu a \; , \qquad
\end{eqnarray}
where $a(x)$ is a(n auxiliary) scalar field, called
{\it PST scalar}, $*H_3$ is defined in (\ref{*H:=}), and
\begin{eqnarray}\label{ivH=}
i_vH_3= {1\over 2}d x^\rho \wedge dx^\nu v^\mu H_{\mu\nu\rho}(x) \; , \qquad  \end{eqnarray} is the contraction
of the three form (\ref{H=dB}) with the vector field $v^\mu =g^{\mu\nu}v_\nu$ dual to the one form $v$ in
(\ref{v=da}). Notice that, by definition,
\begin{eqnarray}\label{v2=pm1}
v^2= v^\mu v_\mu=\pm 1\; . \qquad  \end{eqnarray} As we discuss below (and as we have already mentioned in the
Introduction)  the sign plus    corresponds  to the {\it da-timelike} branch and the sign minus to the {\it
da-spacelike} branch of the PST system.

In the spacetime of nontrivial topology it may be convenient to consider   $da(x)$ as closed but not
exact 1-form, this is to say to consider  $a$  as an angle variable (which implies an equivalence  relation of
the type $a\sim a+2\pi$). This is consistent as $a(x)$ enters the action under derivative so that the constant
shift of its value, $a(x)\mapsto a(x)+const$ is a symmetry of the action.

When working with the differential form representation, it is convenient to use the identities
\begin{eqnarray}\label{v-id}
 F_p\equiv  i_vF_p\wedge v/v^2 + *(i_v*F_p\wedge v)/v^2 \; , \qquad F_6=i_v F_6\wedge v/v^2
 \; ,  \qquad \\ \label{v-idcH}  i_v * {\cal H}_{3} \equiv * ({\cal H}_{3} \wedge  v) \;, \qquad
\end{eqnarray}
\footnote{To have $**=1$ we define $*B_2= -{1\over 4!}dx^\rho_4\wedge ...\wedge dx^\rho_1 {\sqrt{|g|}\over
2}\epsilon_{\rho_1\ldots \rho_4\mu\nu}B^{\mu\nu}$ while keeping the plus sign in the definition $*G_4= {1\over
2}dx^\nu \wedge dx^\mu {\sqrt{|g|}\over 4!}\epsilon_{\mu\nu\rho_1\ldots \rho_4}G^{\rho_1\ldots \rho_4}$. This
provides the plus sign in  (\ref{v-idcH}). } and $F_p\wedge *G_p= G_p\wedge *F_p$. We can also write the first
equation in (\ref{v-id}) as
\begin{eqnarray}\label{v-id-2}
F_p\equiv i_vF_p\wedge {v\over v^2} + F^{(-)}_p\; , \qquad   i_v F^{(-)}=0\; ,
\qquad F^{(-)}_p= {*(i_v*F_p\wedge v)\over v^2}\; . \qquad
\end{eqnarray}
In the case of $da=dt$, $F^{(-)}_p$ is a $p$--form on the spacial slice $M^5_t$ of the spacetime $M^{1+5}$.

\subsubsection{Lagrangian equations from PST action}

We  begin by reviewing the properties of the PST action $\propto \int_{M^6}{\cal L}^{PST}_6 $ in the case of
topologically trivial spacetime.

The variation of ${\cal L}_6^{PST}$ reads
\begin{eqnarray}\label{vcLvH3}
\delta {\cal L}^{PST}_6 =  \pm 2 {\cal G}_2 \wedge da \wedge  \left( \delta H_3 - {1\over 2} d(\delta a)\wedge
{\cal G}_2 \right) \mp H_{3 }\wedge\delta H_3   \;  \qquad
\end{eqnarray}
where the sign $\mp$ in the last term corresponds to $\pm 1=v^2$ and
\begin{eqnarray}\label{cG2:=}
{\cal G}_2:= {i_v({ *{H}_{3} +   {H}_3)}\over \sqrt{\partial a
\partial a}}\; .
\end{eqnarray}
As we are interested in $H_3=dB_2$, or (\ref{H=dB+}) but with  the variation within a  fixed topological class
(\ref{vH=dvB-2}),
\begin{eqnarray}\label{vLHH=}
\delta {\cal L}^{PST}_6  =  2 {\cal G}_{2} \wedge da \wedge  \left( d\delta {B}_2 - {1\over 2}d(\delta a)\wedge
\,{\cal G}_2 \right) \mp  d(H_3 \wedge \delta {B}_2 ) \;.
\end{eqnarray}
For the case of spacetime without boundary,  $\partial M^6=0$, the last term does not contribute into the
variation of the action $S^{PST} \propto \int_{M^6}{\cal L}^{PST}_6$.

Now it is easy to see that the Lagrangian equation of motion for the 2-form potential, which follows from the
PST action, reads
\begin{eqnarray}\label{dG2da=0}
d({\cal G}_{2} \wedge da)=0\; .
\end{eqnarray}
The equation of motion for the PST scalar have the form
\begin{eqnarray}\label{G2dG2da=0}{\cal G}_{2} \wedge d({\cal G}_{2} \wedge da)=0\; .
\end{eqnarray}
Clearly this is satisfied identically due to (\ref{dG2da=0}). This dependence of the $\delta a$ equation makes
transparent the pure gauge (St\"{u}ckelberg) nature of the PST scalar; in other words, it is the Noether
identity which manifests the presence of the gauge symmetry (called PST2 gauge symmetry) with respect to
arbitrary variations $\delta a(x)$  of the PST scalar supplemented by a suitable variation of $B_2$ (the second
term in  (\ref{vB=PST}) below).

But this is not the end of story. Eqs. (\ref{dG2da=0})  can be solved formally with respect to ${\cal G}_{2}$.
In the case of topologically trivial spacetime $M^6$ (with $b_2=0$ and $b_3=0$) the solution is
\begin{eqnarray}\label{cG2da=dp1da}
{\cal G}_{2} \wedge da= -d(\phi_1\wedge da )\; ,
\end{eqnarray}
where $\phi_1=dx^\mu \phi_\mu(x)$ is an arbitrary 1-form.

\subsubsection{Gauge symmetries and branches of the PST system}

Now we come back to Eq. (\ref{vLHH=}) and observe that it also  makes manifest the gauge symmetries which act
on the 2-form potential by
\begin{eqnarray}\label{vB=PST}
\delta {B}_2 = \varphi_1  \wedge da + \delta a \; {\cal G}_{2} \; .
 \end{eqnarray}
Here $\varphi_1 = dx^\mu  \varphi_\mu(x)$ and $\varphi_\mu(x)$ is an arbitrary x-dependent vector function
parametrizing the {\it PST1 gauge symmetry} and  $\delta a=\delta a(x)$ is an arbitrary x-dependent variation
of the PST scalar $a(x)$ (parametrizing the {\it PST2 gauge symmetry}).

Roughly speaking, the PST2 gauge symmetry can be used to gauge $a(x)$ away. However, not all the gauges are
admissible as far as the presence of $\sqrt{\partial a\partial a}$ in the denominator of the Lagrangian, and of
the equations of  motion as they obtained from the variation of the action (Lagrangian equations), put some
topological restrictions on the PST scalar. Neither $a(x)= const$ (which implies $\partial_\mu a=0$) nor
identification of $a(x)$  with the combination of coordinate  parametrizing a light--like direction (for which
$\sqrt{|\partial a\partial a |}=0$) is allowed. However, the configurations where $a(x)$ is identified with
coordinate in timelike or a spacelike direction are not forbidden.

In the case when $\partial_\mu a$ is a timelike vector, $\partial a
\partial a:=\partial_\mu a g^{\mu\nu}
\partial_\nu a>0$, the PST2 gauge symmetry of the PST action can be used to fix the gauge $a(x)=x^0=t$ (or
better $da(x)=dt$).
In the case of the PST action with spacelike $\partial_\mu a$, $\partial a
\partial a <0$, we can use the PST2 gauge symmetry to fix the gauge   $a(x)=x^5$ (better $da(x)=dx^5$). We will
call these {\it da-timelike}  and {\it da-spacelike} branches of the PST system, respectively. These  branches are
disconnected as far as  the configurations $a(x)=t$ and, say, $a(x)=x^5$ cannot be related  by a smooth PST2
transformation\footnote{Indeed, let us  define  one-parametric family of the configurations $a(x)=(1-\alpha
)t+\alpha x^1$, which coincides with $a(x)=t$ and $a(x)=x^1$  for $\alpha=0$ and  $\alpha=1$, respectively. The
fact that  $\alpha =1/2$ representative of this family  corresponds  the prohibited configuration
$a(x)=1/2(t+x^1)$, for which  $\partial a \partial a=0$ and the action and the Lagrangian equations become 
singular, indicates that the PST2 transformation relating the
$a(x)=t$ and $a(x)=x^1$ configurations is inevitably singular. This shows that, even in a topologically trivial
situation, the dynamical system described by the PST action has two disconnected branches, which we call {\it
da-timelike} and {\it da-spacelike} branches of the PST system. }.

\subsubsection{Gauge fixed form of the PST action: HT and PS actions}

Fixing the gauge  $da=dt$   in  {\it da-timelike}  branch of the PST system we arrive at non-manifestly Lorentz-
invariant (non-manifestly diffeomorphism invariant) Henneaux--Teitelboim action
\cite{Henneaux:1987hz,Henneaux:1988gg}
\begin{eqnarray}\label{L6HT:=}
S^{HT}  \propto  \int_{M^6} {\cal L}^{HT}_6\; , \qquad  {\cal L}^{HT}_6= - {\cal G}^{HT}_2\wedge H_3\wedge dt\; ,
\qquad    {\cal G}^{HT}_2=i_0(H_3+*H_3) \; . \quad
\end{eqnarray}
On the other hand, fixing the gauge (say) $da=dx^5$ in  {\it da-spacelike} branch results in the Perry--Schwarz
action \cite{Perry:1996mk}
\begin{eqnarray}\label{L6PS:=}
S^{PS} \propto  \int_{M^6} {\cal L}^{PS}_6\; , \qquad  {\cal L}^{PS}_6= - {\cal G}^{PS}_2\wedge H_3\wedge dx^5 \; ,
\qquad    {\cal G}^{PS}_2=i_5(H_3+*H_3) \; . \quad
\end{eqnarray}

It is easy to write the (counterparts of the) PST1 gauge  symmetry leaving invariant (\ref{L6HT:=}) and
(\ref{L6PS:=}):
\begin{eqnarray}\label{vB=HTsym}
& \delta {B}_2 = \varphi_1  \wedge dt  \;\qquad & \Rightarrow  \qquad \delta S^{HT}=0\; ,\qquad \\
 \label{vB=PSsym}
& \delta {B}_2 = \varphi_1  \wedge dx^5  \qquad & \Rightarrow  \qquad  \delta S^{PS}=0\; .\qquad
 \end{eqnarray}

\subsubsection{Anti-self-duality from Lagrangian equations of motion}

Under the PST1 gauge symmetry $\delta {H}_3 = d(\varphi_1  \wedge da )$ and $\delta {\cal G}_2 \wedge da  = -
d(\varphi_1 ) \wedge da$ so that, making this transformation in (\ref{cG2da=dp1da}) and choosing $\varphi_1 =\phi_1$  we arrive at ${\cal G}_{2} =0 $ which is
equivalent to the anti-self-duality condition (\ref{H=*H}),
\begin{eqnarray}\label{cG2=0=H+*H}
{\cal G}_{2} =0 \qquad \Leftrightarrow \qquad H_3+*H_3=0\; .  \qquad
\end{eqnarray}
In the same manner one can proceed in the case of HT and PS actions (\ref{L6HT:=}) and (\ref{L6PS:=}).

Notice that there exists another, although equivalent,  way of reasonings, which is often used in discussion of
not manifestly invariant HT and PS actions. First one observes  that  the symmetry (\ref{vB=HTsym}) of the HT
action (\ref{L6HT:=}) actually implies that $B_{0i}$ component is absent in it. In other words, ${\cal
L}_6^{HT}$ contains  $B_{ij}=-B_{ji}$ fields only,
 \begin{eqnarray}\label{L6HT:=dB-}
 {\cal L}^{HT}_6= - i_0(dB_2^{(-)}+*(dB_2^{(-)}))\wedge dB_2^{(-)} \wedge dt\; , \qquad    B_2^{(-)}= {1\over
 2}dx^i\wedge dx^jB_{ji}(x)\; .   \qquad
\end{eqnarray}
 Varying with respect to these  fields we find
 \begin{eqnarray}\label{gG2dt=0}
d({\cal G}^{(-)}_{2}\wedge dt)=0\; , \qquad {\cal G}^{(-)}_2= i_0(dB_2^{(-)} +*dB_2^{(-)}) \;  . \qquad
\end{eqnarray}
Notice that  $(dB_2^-)_{ijk}= H_{ijk}:=(dB_2)_{ijk}$ does not feel presence or absence of $B_{0j}$ component so
that  $i_0(*dB_2^-)=i_0(*dB_2)$ and we can write the Lagrangian equations of motion (\ref{gG2dt=0}) in the
following first order form ({\it cf.} (\ref{cG2da=dp1da}))
\begin{eqnarray}\label{cG2dt=dp1dt}
{\cal G}^{(-)}_{2} \wedge dt= d\phi_1\wedge dt \;  \qquad \Leftrightarrow \qquad
(dB_2^{(-)})_{0ij}+(*dB_2)_{0ij}= \partial_i  \phi_j - \partial_j  \phi_i\; .
\end{eqnarray}
This includes $(dB_2^-)_{0ij}= \partial_{0}B_{ij}$  while  the complete $(dB_2)_{0ij}= \partial_{0}B_{ij}- 2
\partial_{[i|}B_{0|j]}$ would involve $B_{0j}$ absent in (\ref{L6HT:=dB-}). The next, final observation is that
we can identify the component $B_{0i}$, absent in the HT action (\ref{L6HT:=}),    with the arbitrary $\phi_i$
appeared in the solution (\ref{cG2dt=dp1dt}),
\begin{eqnarray}\label{B0i=phii}
 B_{0i}=  \phi_i\; ,
\end{eqnarray}
after which (\ref{cG2dt=dp1dt})   acquires the form $(dB_2)_{0ij}+(*dB_2)_{0ij}= 0$ and implies the anti-self
duality equation for the standard field strength
\begin{eqnarray}\label{dB20ij+}
 (dB_2)_{0ij}\vert_{_{B_{0i}=  \phi_i}}+{1\over 3!}\epsilon_{ijklm}(dB_2)^{klm}= 0 \; , \quad \Rightarrow \;
 H_3+*H_3=0\; , \qquad H_3=dB_2\vert_{_{B_{0i}=  \phi_i}}
 \; . \qquad
\end{eqnarray}
It is useful to keep in mind this alternative way of thinking on the derivation of anti-self-duality relation
from the HT action, and similar approach to the PS formulation. However, let us stress that the two forms of HT
action, (\ref{L6HT:=}) and (\ref{L6HT:=dB-}), and two forms of equations of motion, (\ref{cG2dt=dp1dt}) and
(\ref{H=*H}) ((\ref{dB20ij+})), are equivalent because the arbitrary variation of $B_{0i}$ `parametrizes' a
{\it gauge symmetry} of (\ref{L6HT:=}).

Summarizing, as we have reviewed above, in topologically trivial spacetime $M^6=M^{1+5}$, the
anti-self-duality equation (\ref{H=*H}) can be obtained as a gauge fixed version of the Lagrangian equation of
motion (\ref{dG2da=0}) which follows from the PST action (\ref{LHH=}). The main question we address here is
whether this is also the case when the spacetime topology is nontrivial.

\section{PST action and equations of motion in topologically nontrivial $M^{1+5}$}

A topologically nontrivial  spacetime manifold $M^6$ may have some number $b_2$ of closed but not exact 2 forms
$\omega_2^\Lambda$,
\begin{eqnarray}\label{dom2=0}
d\omega_2^\Lambda  =0\; ,\qquad \omega_2^\Lambda \not= d \chi_1^\Lambda  \; , \qquad  \Lambda=1,..., b_2 \, ,
\end{eqnarray}
and also some number $b_3$ of closed but not exact 3-forms $\Omega_3^L$
\begin{eqnarray}\label{dom3=0}
  d\Omega_3^L=0\; , \qquad \Omega_3^L\not=d\Xi_2^L\; , \qquad
L= 1,...,b_3\;  \quad
\end{eqnarray}
(see (\ref{H=dB+})). The number of these closed forms, $b_2$ and $b_3$, determining the  dimensions of the
second and the third de Rahm cohomology groups of the spacetime manifold $M^6$, are called  the second and
third Betti  numbers of  $M^6$,
\begin{eqnarray}\label{b2=}
b_2=\dim {\bb H}^2(M^6)\, , \qquad b_3=\dim {\bb H}^3(M^6)\, .
\end{eqnarray}

The PST action is well defined if $M^6$ admits a nowhere vanishing vector field. As the existence of at least
one such field is necessary condition for a manifold to admit Lorentz-type metric, the PST action always makes
sense in $M^{1+5}$ with nonsingular metric of Lorentzian signature. However, if on $M^{1+5}$  under
consideration such a nowhere vanishing vector field is unique, then both $dt$ and the derivative of PST scalar
$da(x)$ should be  identified with a dual of that; as a result, $da(x)=dt$ and the PST action coincides with the HT one on such a manifold.
The difference between PST and HT action occurs when several (more than one) nowhere
vanishing  vector fields can exist on $M^6=M^{1+5}$. Simplest examples of such spaces are given by  direct products of flat spacetime with an arbitrary internal
manifold,  ${\bb R}^{1+n}\otimes M^{5-n}$ with $5\geq n\geq 1$.

\subsection{First order form of the PST Lagrangian equations  in a topologically nontrivial spacetime}

In a topologically nontrivial  $M^6$, as far as  the Bianchi identities (\ref{dH=0}) are still valid
and we use the variation within fixed topological class $\delta H_3=d\delta B_2$, Eq. (\ref{vH=dvB-2}), the
variation of the Lagrangian form is given by (\ref{vLHH=}), and the Lagrangian equations of motion keep the same
form  (\ref{dG2da=0}),
\begin{eqnarray}\label{dcG2da=0}
d({\cal G}_{2} \wedge da)=0\; . \qquad
\end{eqnarray}
However, when resolving them with respect to ${\cal G}_{2}$ (defined in (\ref{cG2:=})), we first find
\begin{eqnarray}\label{cG2da=gen}
{\cal G}_{2} \wedge da= d\phi_2 +  \tilde{k}_L\Omega_3^L \; . \qquad
\end{eqnarray}
with constant $\tilde{k}_L$ and arbitrary 2-form $\phi_2 $. To proceed, we need to project this differential
form equation into the part which contains $da$ and its complementary, which does not contain $da$. Although
this can be done with generic $da$, the discussion becomes much more transparent if we consider  the case
$da=dt$, in which  Eq. (\ref{cG2da=gen}) acquires the form
\begin{eqnarray}\label{cG2dt=gen}
{\cal G}_{2} \wedge dt= d\phi_2 +  \tilde{k}_L\Omega_3^L \; , \qquad {\cal G}_{2}= i_0 (H_3+*H_3) \; . \qquad
\end{eqnarray}
Let us decompose the forms and differential on the pure spacial and $dt$ dependent parts,
\begin{eqnarray}\label{d=d-+dtdt}
&& d=d^{(-)}+dt\partial_t\; , \qquad d^{(-)}= d\vec{x}\, \vec{\partial}\; , \qquad \\ \label{phi2=-+i0dt} &&
\phi_2  = \phi_2^{(-)}+i_0\phi_2 \wedge dt= \phi_2^{(-)}+\phi_1 \wedge dt \; , \qquad \\ \label{om2=-+i0dt} &&
\omega^\Lambda_2 = \omega_2^{\Lambda\, (-)}+i_0 \omega^\Lambda_2  \wedge dt\; , \qquad
 \\
\label{Om3=-+i0dt} && \Omega^L_3  = \Omega_3^{L\, (-)}+i_0 \Omega^L_3  \wedge dt\; . \qquad
\end{eqnarray}
With splitting (\ref{om2=-+i0dt}) and (\ref{Om3=-+i0dt}), the closure  of basic 2-forms and 3-forms, Eqs. (\ref{dom2=0}) and
(\ref{dom3=0}), imply that the pure spacial part ($^{(-)}$ part) of the closed forms should be `spatially
closed', or $d^{(-)}$--closed forms, and $\partial_t$ derivatives of the spacial ($^{(-)}$) part of the closed
forms should be $d^{(-)}$--exact,
\begin{eqnarray}\label{d-om2-=0}
&& d^{(-)} \omega_2^{\Lambda\, (-)} =0 \; , \qquad \partial_t \omega_2^{\Lambda\, (-)} =  d^{(-)}
i_0\omega_2^{\Lambda} \; , \qquad
\\
\label{d-om3-=0} && d^{(-)} \Omega_3^{L\, (-)} =0 \; , \qquad \partial_t \Omega_3^{L\, (-)} =  d^{(-)}
i_0\Omega_3^{L} \; . \qquad
\end{eqnarray}

Decomposing in the same manner Eq. (\ref{cG2dt=gen}) we find
 \begin{eqnarray}\label{cG2dt=+}
{\cal G}_{2} \wedge dt=  d^{(-)}\phi_1 \wedge dt + (\partial_t \phi_2^{(-)} +  \tilde{k}_Li_0\Omega_3^L) \wedge
dt  \; , \qquad \\ \label{cG2dt=-}  d^{(-)} \phi_2^{(-)} +  \tilde{k}_L \Omega_3^{L\, (-)}=0 \; . \qquad
\end{eqnarray}
Taking into account Eqs. (\ref{d-om3-=0}), we see that (\ref{cG2dt=-}) implies that 2-form $\partial_t
\phi_2^{(-)} +  \tilde{k}_Li_0\Omega_3^L$ is $d^{(-)}$ closed, $
 d^{(-)}(\partial_t \phi_2^{(-)} +  \tilde{k}_Li_0\Omega_3^L)=0\,
$ so that it can be decomposed into a sum of a $d^{(-)}$--exact form and a $d^{(-)}$--closed but not $d^{(-)}$
exact 2--form:
\begin{eqnarray}\label{d-dtom2-+=0} (\partial_t \phi_2^{(-)} +  \tilde{k}_Li_0\Omega_3^L)=
d^{(-)}\check{\phi}^{(-)}_1 + \check{\omega}_2^{(-)}(t,\vec{x})\; , \qquad \begin{cases}
d^{(-)}\check{\omega}_2^{(-)}(t,\vec{x})=0\; , \cr \check{\omega}_2^{(-)}(t,\vec{x})\not=
d^{(-)}\chi_1^{(-)}(t,\vec{x}) \; . \end{cases} \qquad \end{eqnarray}
When substituting this into (\ref{cG2dt=+}), the exact form can be absorbed into the first term, in which
$d^{(-)}$ can be equivalently substituted by $d$,  so that the result reads
\begin{eqnarray}\label{cG2dt=+om-dt}
{\cal G}_{2} \wedge dt= - d\phi_1 \wedge dt +
 \check{\omega}_2^{(-)}(t,\vec{x})\wedge dt  \; , \qquad \begin{cases}
 d^{(-)}\check{\omega}_2^{(-)}(t,\vec{x})=0\; , \cr \check{\omega}_2^{(-)}(t,\vec{x})\not=
 d^{(-)}\chi_1^{(-)}(t,\vec{x}) \; .  \end{cases} \qquad
\end{eqnarray}
A nontrivail topology of $M^6$ may manifest itself in the second term in  right hand side  ({\it r.h.s.}) of this equation.

As one can see from (\ref{d-om2-=0}), a particular case of $d^{(-)}$ closed but not $d^{(-)}$ exact
$\check{\omega}_2^{(-)}(t,\vec{x})$ is provided by pure spacial components of closed but not exact forms
$\omega^\Lambda_2$, possibly multiplied by an arbitrary function of $t$.  In general
\begin{eqnarray}\label{chom2-=}
\check{\omega}_2^{(-)}(t,\vec{x}) = l_\Lambda (t) \omega^{\Lambda (-)}_2 +  \tilde{\omega}_2^{(-)}(t,\vec{x})\, ,
\qquad
  d^{(-)}\tilde{\omega}_2^{(-)}(t,\vec{x})=0 \, ,  \quad \partial_t \tilde{\omega}_2^{(-)}(t,\vec{x})\not=
  d^{(-)} \chi_1^{(-)}  \quad
\end{eqnarray}
with $b_2$ arbitrary functions $l_\Lambda (t)$ of time coordinate only and spatially closed
$\tilde{\omega}_2^{(-)}(t,\vec{x})$ the time derivative of which is not $d^{(-)}$--exact (so that, in
distinction with  $\omega^{\Lambda (-)}_2$,  this is not a spacial part of a closed form in $M^{1+5}$).

In the case of $M^{1+6}= {\bb R}^1\otimes M^5$ all the nontrivial forms have only spacial parts, in particular
$\omega^{\Lambda}_2=\omega^{\Lambda(-)}_2(\vec{x})$, so that all the
$\check{\omega}_2^{(-)}=\check{\omega}_2^{(-)}(\vec{x})$ is decomposed on these, the last term is absent,
$\tilde{\omega}_2^{(-)}=0$, and Eq. (\ref{cG2dt=+om-dt}) reads
\begin{eqnarray}\label{cG2dt=+fom}
{\cal G}_{2} \wedge dt=  d\phi_1 \wedge dt + l_\Lambda (t) \omega^\Lambda_2 \wedge dt  \; .  \qquad
\end{eqnarray}

In this suggestive case, which we will widely use in our discussion below, it becomes especially transparent
that the presence or absence of nontrivial 3-forms $\Omega^L_3$ ($b_3\not=0$ versus $b_3=0$) is not important
when studying the consequence of the Lagrangian PST equation. What does matter is the possible presence of
closed but not exact 2 forms in the spacetime ($b_2\not=0$ versus $b_2=0$) and on its 5 dimensional slices
(spacial slices $M^5_t$ in the case under consideration).

When $da$ is not identified with $dt$, one can also arrive at the counterpart of Eq.  (\ref{cG2dt=+om-dt})
\begin{eqnarray}\label{cG2da=+om-da}
{\cal G}_{2} \wedge da= - d\phi_1 \wedge da + \check{\omega}_2\wedge da\; , \qquad
 d\check{\omega}_2= \check{\omega}_2^{(1)}\wedge d a   \; . \qquad
\end{eqnarray}
Here $\check{\omega}_2^{(1)}$ can be considered as an arbitrary 2-form  which can be identified with
${d\over da}\check{\omega}_2$. However, the consistency conditions of the second equation  in
(\ref{cG2da=+om-da}) require it to obey the equation of the same structure  $d\check{\omega}_2^{(1)} =
\check{\omega}_2^{(2)}\wedge d a$. This chain is continued up to infinity ($d\check{\omega}_2^{(2)} =
\check{\omega}_2^{(3)}\wedge d a$, $\ldots $, $d\check{\omega}_2^{(n)} =  \check{\omega}_2^{(n+1)}\wedge da$,
$\ldots $) and represents a  counterpart of $d^{(-)}\check{\omega}_2=0$ condition in (\ref{cG2dt=+om-dt})
(actually it can be equivalently written in this form provided $d^{(-)}\check{\omega}_2=d\check{\omega}_2\mp
i_vd\check{\omega}_2\wedge v$ with $v\propto da$ (\ref{v=da})).

In some cases (for some $M^6$) we can write Eq. (\ref{cG2da=+om-da}) in the form similar to
(\ref{cG2dt=+fom}),
\begin{eqnarray}\label{cG2da=+fom}
{\cal G}_{2} \wedge da= d\phi_1 \wedge da + l_\Lambda (a(x)) \omega^\Lambda_2 \wedge da(x) \; , \qquad \Lambda
=1,..., {b}_2 \; ,   \qquad
\end{eqnarray}
In addition to  the arbitrary 1-form $\phi_1$, the {\it r.h.s.} of this equation  contains a
topological contributions determined by  ${b}_2$ arbitrary functions of the PST scalar, $l_\Lambda
(a)=l_\Lambda (a(x))$. As in the case of topologically trivial $M^6$, the first term can be gauged away using
the PST1 gauge symmetry, but the fate of the second term, containing topological contribution, have to be
studied.

\subsection{`Semilocal symmetry' of the PST action in topologically nontrivial $M^6$}

\label{sec=semi}

Coming back to the variation of the Lagrangian form, Eq. (\ref{vLHH=}), we find that in a topologically
nontrivial spacetime $M^6=M^{1+5}$ it vanishes under the transformations of the PST2 gauge symmetry $\delta
B_2= \delta a(x)\,  {\cal G}_2$ (which are the same as in the topologically trivial case) and also
 under the variations $\delta H_3=d\delta B_2$ which obey
\begin{eqnarray}\label{dadvB=0}
da \wedge d\delta B_2 \equiv d(\delta B_2  \wedge da)=0 \; .
\end{eqnarray}

Actually this  equation for $\delta B_2^{(-)}:= \delta B_2  \mp i_v \delta B_2 \wedge v$ (with $v \propto da$,
Eq. (\ref{v=da})) have the same structure as  Eq. (\ref{dcG2da=0}) for ${\cal G}_2\propto i_v (H_3+*H_3)$.
Hence, using the results of previous subsection,  we can immediately write its general solution for $\delta
B_2$.  Combining it with the above mentioned PST2 transformations (and ignoring the conventional gauge symmetry
$\delta B_2=d\alpha_1$), we find
\begin{eqnarray}\label{vB=PST+TopG}
\delta {B}_2 = \varphi_1  \wedge da + \delta a \; {\cal G}_{2} +
 \check{\varphi}_2\; , \qquad d\check{\varphi}_2= \check{\varphi}^{(1)}_2 \wedge da \; , \qquad
 \end{eqnarray}
which, at least in some particular cases
(see above) can be written as
\begin{eqnarray}\label{vB=PST+Top}
\delta {B}_2 = \varphi_1  \wedge da + \delta a \; {\cal G}_{2} +
 \omega_2^\Lambda \; f_\Lambda(a(x))\; \qquad
 \end{eqnarray}
(see (\ref{dom2=0}) for properties of $\omega_2^\Lambda$).

In addition to an arbitrary x-dependent one form variation  $\varphi_1 = dx^\mu  \varphi_\mu(x)$, and
 an arbitrary variation of the PST scalar $\delta a(x)$ (parametrizing the PST1 and PST2 gauge symmetries,
 respectively), this contains $b_2$ arbitrary functions  of the PST scalar field $a=a(x)\;$,
 $f_\Lambda(a)=f_\Lambda(a(x))$.
These parametrize   transformations of {\it semilocal symmetry}.

In general case the semi-local symmetry is described by two-form $\check{\varphi}_2$ in (\ref{vB=PST+TopG})
which obeys $d\check{\varphi}_2=  \check{\varphi}^{(1)}_2 \wedge da$; as such a description seems to be less
transparent, it is very useful to keep in mind the particular case described above.

To gain more  comprehension of the properties of our system,  let us discuss the configurations with $da=dt$ of the $da$-timelike branch of the PST system, or  the case of HT action. Then  (\ref{vB=PST+TopG}) acquires the
form
\begin{eqnarray}\label{vB=PSTdt+TopG}
\delta {B}_2 = \varphi_1  \wedge dt +
 \check{\varphi}_2^{(-)}\; , \qquad d^{(-)}\check{\varphi}_2^{(-)}= 0 \; , \qquad
 \end{eqnarray}
 and (\ref{vB=PST+Top}), valid in particular cases (including $M^6= {\bb R}^1\otimes M^5\;$
  \footnote{The semi-local symmetry of the HT action in such spacetime was observed in
  \cite{Bekaert:1998yp}.}), reads
\begin{eqnarray}\label{vB=PSTdt+Top}
\delta {B}_2 = \varphi_1  \wedge dt +
 \omega_2^\Lambda \; f_\Lambda(t)\; . \qquad
 \end{eqnarray}
We see that in these cases the semilocal symmetry  transformations are parametrized by $b_2$ functions of time
variables only, $f_\Lambda(t)$. In general case of topologically nontrivial spacetime $M^{1+5}$, the
parameters of the semi-local symmetry of HT action  are hidden inside a $t$-dependent  $d^{(-)}$--closed but
not $d^{(-)}$--exact 2-form,  $\check{\varphi}_2^{(-)}(t,\vec{x})$ in  (\ref{vB=PSTdt+TopG}).

The symmetry transformations of the field strength  $H_3=dB_2$ read
\begin{eqnarray}\label{vH3=PSTdt+TopG}
\delta {H}_3 = d(\varphi_1  \wedge dt ) -
  \partial_t\check{\varphi}_2^{(-)}\wedge dt\; , \qquad d^{(-)}\check{\varphi}_2^{(-)}(t,\vec{x})= 0 \; ,
  \qquad
\end{eqnarray}
and, in the particular case,
\begin{eqnarray}\label{vH3=PSTdt+Top}
\delta {H}_3 = d(\varphi_1  \wedge dt ) +
  \dot{f}_\Lambda (t)\; \omega_2^\Lambda \wedge dt\; ,
\end{eqnarray}
 where $\dot{f}_\Lambda (t):= {df_\Lambda (t)\over dt}$. As it is easily to see, $i_0 *\delta {H}_3= i_0
 \delta *{H}_3=0$, so that
\begin{eqnarray}\label{vG2dt=PST+top}
\delta {\cal G}_2 \wedge dt  = - d(\varphi_1 ) \wedge dt  +  \partial_t\check{\varphi}_2^{(-)}\wedge dt\; ,
\qquad d^{(-)}\check{\varphi}_2^{(-)}(t,\vec{x})= 0 \; .  \qquad
\end{eqnarray}
Eq. (\ref{vG2dt=PST+top}) implies that,  as in the topologically trivial case, we can gauge away  $\phi_1$ in
(\ref{cG2dt=+om-dt}) using the second PST symmetry with the one-form parameter $\varphi_1 $. Furthermore, the
second term in (\ref{cG2dt=+om-dt}) can be also removed by the 'semi-local symmetry' if we choose
$\check{\varphi}_2^{(-)}(t,\vec{x})$ to be a solution of $\partial_t\check{\varphi}_2^{(-)}= -
 \check{\omega}_2^{(-)}(t,\vec{x})$.

In particular cases when $\check{\varphi}_2^{(-)}(t,\vec{x})$ and $\check{\omega}_2^{(-)}(t,\vec{x})$ can be
expressed as a linear combination of $\omega_2^{(-)\Lambda}(t,\vec{x})$, the above arguments can be formulated
in a more transparent  manner:  the
{\it r.h.s.} (\ref{cG2dt=+fom}) can be removed by a semi-local symmetry  provided the $b_2$ functions in
(\ref{vB=PSTdt+Top}) (and (\ref{vH3=PSTdt+Top})) are chosen such that $\dot{f}_\Lambda (t)= - l_\Lambda (t)$
holds.

Similar statement is true in the generic case of (both branches of) the PST action with an arbitrary (nowhere
vanishing) $da$. Let us begin from the PST system in particular type of spacetime where the general solution of the Lagrangian PST equation
(\ref{dcG2da=0}) can be written in the form of (\ref{cG2da=+fom}). The {\it r.h.s.} of this equation can
be removed by the standard PST gauge symmetries and the semi-local symmetry $\delta B_2= \omega_2^\Lambda \;
f_\Lambda (a, \delta )$, Eq. (\ref{vB=PST+Top}), with  $a$-dependent variation  $f_\Lambda (a)$ obeying $
f'_\Lambda (a):= {d\over da}f_\Lambda (a)= - l_\Lambda (a)$.

In general case the {\it r.h.s.} of the first order form of the PST Lagrangian equation, Eq.
(\ref{cG2da=+om-da}), contains 2-form $\check{\omega}_2$ obeying $d\check{\omega}_2=\check{\omega}^{(1)}_2\wedge
da$, the {\it r.h.s.} of (\ref{cG2da=+om-da}) can be also removed by the standard PST gauge symmetries and the
semi-local symmetry, $\delta B_2= \check{\varphi}_2$ with $d\check{\varphi}_2= \check{\varphi}_2^{(1)}\wedge
da$ in (\ref{vB=PST+TopG}), provided we choose $\check{\varphi}_2^{(1)}= -\check{\omega}_2$.

Then, {\it if the semilocal symmetry
is a gauge symmetry}, the above statements imply that Eq. (\ref{cG2da=+om-da}) (Eq.
(\ref{cG2da=+fom})) is gauge equivalent to  ${\cal G}_2=0 $ which, in its turn, is equivalent to the usual
anti--self-duality equation $H_3+*H_3=0$, Eq. (\ref{cG2=0=H+*H}).

As we will show in  sec. \ref{sec=Noether} this is the case in the {\it da-timelike} branch of the PST
system, while in the  {\it da-spacelike branch}  the above semilocal symmetry is an infinitely dimensional
global symmetry, similar to d=2 conformal symmetry.

To make the above  statement intuitively clear, we just mention that  the gauge nature of semilocal
symmetry in the $da$-timelike PST system is suggested by observation that in it, after gauge fixing the PST2
symmetry by $da=dt$, the parameters of the semilocal symmetry (\ref{vB=PST+TopG}) (or (\ref{vB=PST+Top})) can
be collected in functions of time variables; in a particular case these are $b_2$ functions  $f_\Lambda (t)$
(see (\ref{vB=PSTdt+Top})).  Indeed,  the t-dependence  is characteristic for parametric functions of gauge
symmetries of  one-dimensional systems,  and, in a $d \geq 2$ field theory, the dependence on  spacial
coordinates can be considered as a kind of index, although continuous. In contrast, in the {\it da-spacelike}
branch the parameters of semilocal symmetry can be collected (after gauge fixing the PST2 symmetry by, say,
$da=dx^5$) in functions of one of the spacial coordinates, which suggests an  infinite dimensional global
symmetry nature of semilocal symmetry in this case.

\subsection{On Noether current for semilocal symmetry}
\label{sec=Noether}

A formal way to distinguish a gauge symmetry from an infinite dimensional rigid symmetry, a typical example of
which is provided by the 2d conformal symmetry, is to calculate the Noether current $J^\mu$ and Noether charge
$Q=\int d^{D-1}x J^0$. For the gauge  symmetry this latter is identically  equal to zero, $Q=\int d^{D-1}x
J^0=0$, while for the rigid symmetry this is not the case (see \cite{Sevrin:2013nca} and refs. therein).

In this section we address the question of whether the semilocal symmetry of the PST action in a topologically
nontrivial spacetime,  which in some particular cases of $M^6$  with $b_2\not= 0$ have the form of $\delta B_2=
\omega_2^\Lambda \; f_\Lambda(a, \delta )$  in (\ref{vB=PST+Top}), is the gauge  symmetry.

To streamline the discussion, we can consider the non-manifestly Lorentz invariant Henneaux-Teitelboim (HT)
and Perry--Schwarz (PS) actions, Eqs. (\ref{L6HT:=}) and (\ref{L6PS:=}), which  can be obtained by gauge fixing from the PST action in its {\it da-timelike} and {\it da-specelike} branches, respectively.

Furthermore, we will begin by discussing these actions in particular classes of topologically nontrivial
spacetimes, including ${\bb R}^1\otimes M^5$ and $M^{1+4}\otimes {\bb R}^1$, for which the semi-local symmetry of these actions can be expressed in terms of $b_2$ functions of only time coordinate and of only one spacial coordinate
respectively, so that
\begin{eqnarray}\label{PST1-HT}
\delta S^{HT}=0 \;  \quad &\Leftarrow & \quad \delta B_2 =d\alpha_1+ \varphi_1 \wedge dt + f_\Lambda(t)
\omega_2^\Lambda  \; . \quad \\ \label{PST1-PS} \delta S^{PS}=0 \;  \quad &\Leftarrow &\quad \delta B_2
=d\alpha_1+ \varphi_1 \wedge dx^5 + f_\Lambda(x^5) \omega_2^\Lambda  \; . \quad
\end{eqnarray}
Let us recall that $\omega^\Lambda$, $\Lambda=1,..., b_2$  are closed but not exact 2--forms (\ref{dom2=0})
which  provide a basis  of ${\bb H}^2(M^6)$ which, in these particular cases, is equal to  ${\bb H}^2(M^5)$ and
${\bb H}^2(M^{1+4})$, respectively.

In the case of HT action for chiral bosons in ${\bb R}^1\otimes M^5$, we can write the semi-local symmetry
transformations of (\ref{PST1-HT}) in the form \begin{eqnarray}\label{semi-HT} \delta B_2= \left( f_\Lambda(0)+
t f'_\Lambda (0)+...+t^n{f^{(n)}_\Lambda(0)\over n!} +...\right) \; \omega_2^\Lambda\;
\end{eqnarray}
and identify (the infinite set of) their  parameters  with ${f^{(n)}_\Lambda(0)\over n!}$ for $n=0,1,...\;$. Then
the 5-forms dual to the Noether currents for these symmetries,
 \begin{eqnarray} \label{*J1=d5xeJ}
*J_1^{(n)\Lambda }= {1\over 5!} dx^{\mu_5} \wedge \ldots \wedge dx^{\mu_1} \epsilon_{\mu_1\ldots \mu_5 \mu}
J^{\mu (n)\Lambda}\; . \qquad
\end{eqnarray}
read   (see Appendix B for the proof of the First Noether Theorem in our notation)
\begin{eqnarray}\label{NCu-HT}
*J_1^{(n)\Lambda }= t^n dt\wedge {\cal G}_2 \wedge \omega_2^\Lambda\; . \quad
\end{eqnarray}
This form is closed on the mass shell, $d*J_1^{(n)\Lambda }=0$, which is tantamount to state the current
conservation $\partial_\mu J^{\mu (n)\Lambda}=0$. Moreover, one can observe that the time component of the
Noether current dual to the 5-form (\ref{NCu-HT}) vanishes, $J^{0 (n)\Lambda }=0$ ($J^{\mu (n)\Lambda
}=\delta^\mu_i J^{i(n)\Lambda }$). Then, as far as the  Noether charge is defined as an integral over the space
of the timelike component of the Noether current, it also vanishes, \begin{eqnarray}\label{Qn=intJ0}
 Q^{(n)\Lambda }:=\int d^5x J^{0(n)\Lambda }=0\; , \quad
\end{eqnarray} as it should be for the case of gauge symmetry.

This allows us to conclude that the semi-local symmetry (\ref{PST1-HT}) is the gauge symmetry of the HT action
(\ref{L6HT:=}) in a spacetime $M^6$ of nontrivial topology with $b_2\not= 0$.

Such a conclusion does not follow in the case of semi-local symmetry (\ref{PST1-PS}) of the Perry--Schwarz
action (\ref{L6PS:=}) on $M^{1+4}\otimes {\bb R}^1$,
\begin{eqnarray}
\label{semi-PS} \delta B_2= \left( f_\Lambda(0)+ (x^5) f'_\Lambda (0)+...+(x^5)^n{f^{(n)}_\Lambda(0)\over n!}
+...\right)\; \omega_2^\Lambda .
 \quad
\end{eqnarray}
Its  Noether current reads
\begin{eqnarray}\label{NCu-PS}
*J_1^{(n)\Lambda }= (x^5)^n dx^5\wedge {\cal G}_2 \wedge \omega_2^\Lambda\; , \quad
\end{eqnarray}
so that in this case $J^{5(n)\Lambda }=0$, but $J^{0(n)\Lambda }$ is generically nonvanishing. Thus the {\it
standard} Noether charge is generically nonzero,  $ Q^{(n)\Lambda }:=\int d^5x J^{0(n)\Lambda }\not=0$, which
indicates that the semi-local symmetry (\ref{PST1-PS}) of (\ref{L6PS:=}) is infinite dimensional rigid
symmetry.

The above statement is true also in the case of HT, PS and PST actions in generic $M^6$ (allowing for the
existence of nowhere vanishing vector field(s)), when the semi-local symmetry is described by Eq.
(\ref{vB=PST+TopG}) rather than  (\ref{vB=PST+Top}). To make this transparent let us first return to the above
particular case and introduce generated functionals for the Noether currents and Noether charges,
\begin{eqnarray}\label{NCuf-HT}
& *J_1^{\Lambda }[f(t)]= \sum\limits_{n=0}^{\infty} {f^{(n)}(0)\over n!}*J_1^{(n)\Lambda }= f(t) dt\wedge {\cal
G}_2 \wedge \omega_2^\Lambda\; , \qquad \nonumber \\ & Q^{\Lambda }[f(t)]=\int d^5x J^{0 \Lambda }[f(t)] \equiv
\int_{M^5}  *J_1^{\Lambda }[f(t)] =0 , \qquad \nonumber \\ & Q[f_\Lambda(t) \omega^\Lambda_2] =
\sum\limits_{\Lambda} Q^{\Lambda }[f_\Lambda(t)]  =0 , \qquad
 \\ \label{NCuf-PS}
& *J_1^{\Lambda }[f(x^5)]= \sum\limits_{n=0}^{\infty} {f^{(n)}(0)\over n!}*J_1^{(n)\Lambda }= f(x^5) dx^5\wedge
{\cal G}_2 \wedge \omega_2^\Lambda\; , \qquad \nonumber \\ & Q^{\Lambda }[f(t)]=\int d^5x J^{0 \Lambda }[f(t)]
\equiv \int_{M^5}  *J_1^{\Lambda }[f(t)] \not=0 , \qquad \nonumber \\ & Q[f_\Lambda(x^5) \omega^\Lambda_2] =
\sum\limits_{\Lambda} Q^{\Lambda }[f_\Lambda(x^5)] \not=0 . \qquad
\end{eqnarray}

In general the generating function for the Noether changes  of semi-local symmetry (\ref{vB=PST+TopG}) can be
written as
\begin{eqnarray}\label{NCuG-HT}
& Q[\check{\varphi}_2] = \int_{M^5}  da(x)\wedge {\cal G}_2 \wedge \check{\varphi}_2\; , \qquad d
\check{\varphi}_2= \check{\varphi}^{(1)}_2\wedge da , \qquad
\end{eqnarray}
where  $M^5$ can be defined as a constant $t$ slice of $M^{1+5}$. In the $da$--timelike branch of the dynamical
system described by the PST action we can use the PST2 symmetry to fix the gauge $da=dt$ in which it is
immediate to see that $Q[\check{\varphi}_2] =0$. This is not the case for the $da$--spacelike branches of PST
system, where we can rather fix the gauge $da=dx^5$ in which, generically, $Q[\check{\varphi}_2] \not=0$.

Hence the {\it semi-local symmetry is a gauge symmetry in the  $da$--timelike branch} of the dynamical system
described by the PST action {\it and  an infinite dimensional rigid symmetry in the $da$-spacelike branch of the PST
system}.

\subsubsection{ A speculation on possible alternative}

\label{spec}

It is tempting, following the spirit of recent \cite{Hull:2014cxa} (devoted to the Euclidean 5d SYM description
of the mysterious non-Abelian 6d $(2,0)$ superconformal theory), to speculate on possible alternative
canonical formalism allowing to treat the semilocal symmetry of PS action as a gauge symmetry.

Indeed, the above observation that the Noether current of semi-local symmetry of the PS action obeys
$J^{5(n)\Lambda }=0$ (in particular cases, and $i_5* J_1[[\check{\varphi}_2]]=0$ in the general case) implies
vanishing of 'pseudo-charge' constructed in the same way as Noether charge but with interchanging the role of
time $x^0=t$ and the special space direction $x^5$,
\begin{eqnarray}\label{Qn=intJ5}
\tilde{Q}^{(n)\Lambda }:=\int dtdx^1dx^2dx^3dx^4  J^{5(n) \Lambda }=0\; . \quad
\end{eqnarray}
Then, following \cite{Hull:2014cxa}, one can build formally the  canonical formalism based on using $x^5$ instead of time direction $x^0=t$. In its frame the semilocal symmetry  of  the PS
action can be treated as a gauge symmetry and used to obtain the (anti-)self-duality equation. A necessary
condition for the above treatment of PS action and of the  $da$-spacelike branch of the PST system in a
topologically nontrivial spacetime with $b_2\not=0$ is that this $M^6$ allows for the  existence of a nowhere
vanishing spacelike vector field, which can be associated with $x^5$ direction (besides the timelike nowhere
vanishing vector field which is strictly necessary to have the metric of Lorentz signature). If so, one could
state that also $da$-spacelike branch  of the covariant PST action, like $da$-timelike one, can be used to
obtain the anti--self duality equation, provided certain alternative  canonical formalism is used. The authors
of \cite{Hull:2014cxa} noticed that different choices of the basic variables of canonical formalism should
correspond to restriction to a special field configurations for which the physically relevant integrals of the
field variables (including  the ones defining charges) do not diverge.

In this paper we will not elaborate the idea of alternative canonical formalism  and keep the  conservative
point of view according to which the semilocal symmetry of the PS action and of the $da$-spacelike branch of the
PST system is an infinite dimensional rigid symmetry.

\subsection{(Anti-)self-duality equation from  PST action  in topologically nontrivial  $M^{6}$}
\label{b2not=0sec}

Thus, as we have shown above, in the topologically nontrivial space--time $M^{1+5}$  the PST action produces
the dynamical equations of motion (\ref{dG2da=0}),  which are equivalent to (\ref{cG2da=+om-da}). In particular
cases of topologically nontrivial spacetimes this first order form of the Lagrangian equations  can be written
in a more transparent manner (\ref{cG2da=+fom}),
\begin{eqnarray}\label{G2da=PST+top2}
{\cal G}_{2} \wedge da := i_v(H_3+*H_3)\wedge v= l_\Lambda (a) \, \omega^\Lambda \wedge da\; ,
\end{eqnarray}
in which the {\it r.h.s.} is expressed in terms of  $b_2$  arbitrary functions of the PST scalar  $l_\Lambda
=l_\Lambda (a(x))$.  Below we will carry the discussion in terms of this particular case and give the generic form of the equations in parenthesis. (See  (\ref{cG2da=+om-da}) for the first order form of the PST Lagrangian equations in a generic spacetime).

On the other hand the PST action possesses, in addition to  the PST1 and PST2 gauge symmetries, also  the
semi-local symmetry which  in these particular spacetimes, is parametrized by $b_2$ arbitrary
functions of the PST scalar $f_\Lambda(a(x))$ in (\ref{vB=PST+Top}),
 \begin{eqnarray}\label{vB=PST+Top2}
\delta {B}_2 = \varphi_1  \wedge da + \delta a \; {\cal G}_{2} +
 \omega_2^\Lambda \; f_\Lambda(a(x))\;
 \end{eqnarray}
 ((\ref{vB=PST+TopG}) in the generic case).

 \subsubsection*{3.4.1. $da$-timelike branch}
In the  {\it da-timelike branch} of the PST system, described by the PST action with timelike $\partial_\mu a$,
all these are  gauge symmetries and one can use them to gauge away the {\it r.h.s.} of (\ref{G2da=PST+top2}),
thus arriving at ${\cal G}_{2} \wedge da := i_v(H_3+*H_3)\wedge v=0$ which is equivalent to the
anti-self-duality (\ref{cG2=0=H+*H}),
\begin{eqnarray}\label{H+tH=0}
{H}_3+ *{H}_3=0\;  \qquad
\end{eqnarray}
for the 'original' field strength (\ref{H=dB+}) entering the action (\ref{LHH=}).

 \subsubsection*{3.4.2. $da$-spacelike branch}

As we have already discussed in sec. \ref{spec},  to treat similarly the {\it $da$--spacelike} branch of the PST
system, as well as the PS system appearing as the gauge fixed, $da=dx^5$ version of this branch, one might try,
following the spirit of \cite{Hull:2014cxa} to develop an alternative canonical formalism using  $x^5$ variable
instead of time.

In this paper we will follow a more conservative approach keeping the standard relation of  canonical formalism
with the timelike direction of $M^{1+5}$. Then, the semi-local symmetry of the  {\it da-spacelike branch} of the
PST system  is an infinite dimensional global symmetry, which cannot be used to fix a gauge, and the Lagrangian
equation (\ref{dG2da=0}) is gauge equivalent to
\begin{eqnarray}\label{G2da=PST+top3}
{\cal G}_{2} \wedge da := i_v(H_3+*H_3)\wedge v= l_\Lambda (a) \, \omega^\Lambda \wedge da
\end{eqnarray}
{\it with arbitrary functions} of the PST scalar $l_\Lambda (a)$ (to (\ref{cG2da=+om-da}) with $\phi_1=0$  in the general case).

Thus the {\it standard} anti-self-duality equation (\ref{H+tH=0}) {\it for the field strength $H_3$ entering
the action} cannot be reproduced in the  {\it da-spacelike branch}. However, let us observe that we can remove
the additional topological contributions by redefining the field strength as
\begin{eqnarray}\label{tH=H-lomda1}
\tilde{H}_3= H_3 - l_\Lambda (a(x)) \, \omega^\Lambda \wedge da(x) \;  \qquad
\end{eqnarray}
(or, in the general case,
\begin{eqnarray}\label{tH=H-lomdaG}
\tilde{H}_3= H_3 - \check{\omega}_2 \wedge da(x) \;  , \qquad d\check{\omega}_2 = \check{\omega}_2^{(1)} \wedge
da(x)\; , \qquad
 \end{eqnarray}
with $\check{\omega}_2$ being arbitrary in all other respects). Indeed, due to (\ref{G2da=PST+top3}) (or
(\ref{cG2da=+om-da})), such $\tilde{H}_3$ obeys, besides the standard Bianchi identities $d\tilde{H}_3=0$, also
$i_v (\tilde{H}_3+ *\tilde{H}_3)=0$ and hence
\begin{eqnarray}\label{tH+*tH=0}
\tilde{H}_3+ *\tilde{H}_3=0\; . \qquad
\end{eqnarray}

Actually, when we can consider  $da(x)$ (and hence $da(x) l_\Lambda (a)$ in (\ref{tH=H-lomda1})) as an exact
form, we can write the redefinition of the field strength as a redefinition of the potential,
\begin{eqnarray}\label{dBsimdtB}
& {H}_3 =dB_2 \; , \quad \tilde{H}_3=d\tilde{B}_2\; , \qquad  \tilde{B}_2 = B_2 + \omega_2^\Lambda
\tilde{f}_\Lambda (a)+d\alpha_1  \; , \qquad {d\over da}\tilde{f} _\Lambda (a)= l_\Lambda (a)  \quad
\end{eqnarray}
($\tilde{B}_2 = B_2 + \beta_2$ with $d\beta_2= \check{\omega}_2\wedge da$ in general case).

This allows us to state that, in a topologically nontrivial spacetime with $b_2\not=0$, and more generally in
the spacetime which allows for an existence of a nontrivial solution ($\check{\omega}_2\not= d\chi_1 +
i_vd\chi_1\wedge v$) of the equation $d \check{\omega}_2= \check{\omega}^{(1)}_2\wedge da$, the $da$-spacelike
branch of the PST system, as well as the non-manifestly covariant PS action, can be used to produce the
(anti-)self-duality equation  only for the redefined field strength $\tilde{H}_3$  (\ref{tH+*tH=0}) or
redefined potential  $\tilde{B}_2$ (\ref{dBsimdtB}). In particular cases where the redefinition has an
especially transparent form, it involves $b_2$ arbitrary functions of the PST scalar (of a special spacial
coordinate $x^5$ in the case of PS action) which can be associated  with the set of parameters of semilocal
symmetry. However, as far as this symmetry is an infinite dimensional global symmetry rather than a gauge
symmetry,   the original and redefined  potentials, $B_2$ and $\tilde{B}_2$, cannot be considered as equivalent in this case.

Thus the topology makes difference between $da$-timelike and $da$-spacelike branches of the PST system as well as
between non-manifestly invariant HT and PS actions. The timelike branch of the PST system and its gauge fixed
version described by HT action, become preferable as they produce the (anti-)self--duality equation for the
original field strength (field strength of the original potential), which enters the action, also in spacetime
with $b_2\not=0$ and more generally in spacetime allowing for existence of a nontrivial solution
($\check{\omega}_2\not= d\chi_1 -i_vd\chi_1\wedge v$) of the equation $d \check{\omega}_2=
\check{\omega}^{(1)}_2\wedge da$.

 \subsubsection*{3.4.3. On spacetime of Euclidean signature $M^{6}=M^{6+0}$ }

In the case of topologically nontrivial spacetime of Euclidean signature, $M^{6}=M^{6+0}$, the main problem is
that, to obey $*^2=1$, the Hodge duality operation is to be imaginary, $(*H_3)^*= -*(H_3)^*$ so that the
(anti-)self--dual gauge field should be complex,  $(H_3)^* \not= H_3$. However, if we allow for complex fields
and non-Hermitian actions (this is widely used {\it e.g.} in pure spinor approach to quantum superstring
\cite{Berkovits:2000fe,Berkovits:2004px}; see also \cite{Belov:2006jd}), then  the only requirement for the
spacetime $M^{6}=M^{6+0}$ to define (a complex) PST action is to allow for a nowhere vanishing vector fields.
Then if such field is unique, the only possibility is to identify $da$ and $dt$ with it(s dual) and the PST
action automatically reduces to HT one.

In $M^{6}=M^{6+0}$ of Euclidean signature with several nowhere vanishing vector fields all the allowed choices
of $da$ can be related by a nonsingular PST2 gauge transformations so that no separations on the branches in
(complex) PST action occurs and the (complex)  HT and PS actions are gauge equivalent.
In some particular cases, including $M^{6}$ with $b_2(M^6)\not= 0$, the additional contributions to the {\it r.h.s} of the first order form of the Lagrangian equation which follows from the PST action do appear; however the PST action possesses the semi-local symmetry, and this can be treated as gauge symmetry. Hence the (complex) PST action in such an $M^{6}=M^{6+0}$ produces the Lagrangian equations which are gauge equivalent to the  anti-self-duality conditions (\ref{cG2=0=H+*H}) for the original (complex) gauge field strength which enters the action.

\subsection{Summarizing the case of chiral 2-form gauge potential in 6 dimensions}

Thus we have shown that the PST action for the 6D 2-form potential  can be used to obtain the self-duality
equations also in the topologically nontrivial spacetime.

Interestingly enough, the $da$-timelike and $da$-spacelike branches of the PST system  become nonequivalent
 in the spacetime $M^{1+5}$ with
$b_2=dim\,  {\bb H}^2(M^{1+5})\not= 0 $ or, more generally if $M^{1+5}$ allows for a nontrivial solution
($\check{\omega}_2\not= d\chi_1\mp i_{v}d\chi_1\wedge v$) of the equation $d \check{\omega}_2=
\check{\omega}^{(1)}_2\wedge da$ with some closed 1--form $da=da(x)$ ($v\propto da$).  Also the non-manifestly
invariant HT and PS actions, which can be obtained from these branches of PST system by gauge fixing of PST2
symmetry, are non-equivalent in such spacetimes.  In both cases the Lagrangian equations are equivalent to the
(anti-)self-duality conditions for redefined field strength and the redefinition can be identified with some
semi-local symmetry of the PST action. However, the  $da$-timelike PST system is preferable as in it the
semi-local symmetry is a gauge symmetry so that the redefined field strength is gauge equivalent to the
original one which enters the action. Thus, in this $da$-timelike branch of PST system  (and in the HT action) the Lagrangian equations are equivalent to the anti-self-duality equation for the original field strength entering the PST action.

Similar situation occurs for the chiral $2l$-form gauge field in topologically nontrivial spacetime of
$D=4l+2$ dimensions and, with a minimal modification, also for any even $D$. But before describing this general
case, we turn to the simplest D=2 chiral boson system, in which a semi-local symmetry occurs also in PST action
in the topologically trivial D=2 spacetime.

\section{Prototype of the topological gauge symmetry in 2D PST action for chiral bosons}

The simplest, but also special case of theories of self-dual and anti-self-dual tensor fields
is the theory of chiral boson in two dimensional spacetime. In a topologically trivial spacetime one can fix the conformal gauge, where the chiral
(anti-chiral) bosons are represented by functions of only $t-x $ (only  $t+x$)\footnote{ In this section we use
the notation $x^m=(x^0,x^1)=(t,x)$, so that $f(x)$ denotes the function of one rather than of two
coordinates.}. In terms of differential forms, one defines the 2d Hodge star operation by
\begin{eqnarray}\label{Hodge*=2D}
&& {} *d\phi = *(dx^m\partial_m\Phi)=dx^m \sqrt{|g|}\epsilon_{mn}g^{nk} \partial_k\phi\; , \qquad  \nonumber \\
&& {} *1= d^2x \sqrt{|g|}:= {1\over 2}dx^m\wedge dx^n \epsilon_{nm} \sqrt{|g|}, \qquad {} **=I \;  , \qquad
 \end{eqnarray}
with $\epsilon_{mn}=- \epsilon_{nm}$, $\epsilon^{01}=- \epsilon_{01}=1$, and writes the chirality condition
\begin{eqnarray}\label{dtp+dxp=0}
(\partial_t+\partial_x) \phi =0\;  \qquad
 \end{eqnarray}
as 2d anti-self-duality equation,
\begin{eqnarray}\label{d+*d=0}
d\phi + *d\phi =0\; . \qquad
 \end{eqnarray}
The Lagrangian 2-form of 2D PST  action can be written as
\begin{eqnarray}\label{cL}
{\cal L}_2=i_v(d\phi + *d\phi)  \, d\phi \wedge v = {1\over 2v^2}d\phi  \wedge *d\phi
 - {1\over 2} i_v(d\phi + *d\phi) \; * i_v(d\phi + *d\phi)
\; . \qquad
 \end{eqnarray}
 For completeness of this section let us recall that $v=dx^m v_m$ is defined in (\ref{v=da}), $v^2=\pm 1$, and
 $i_v$ -- by $i_v d\phi=v^m\partial_m\phi$ and by the 2-form counterpart  of (\ref{ivH=}). Denoting
$F=d\phi$, we can write the variation of the Lagrangian form  (\ref{cL}) as
\begin{eqnarray}\label{vcL}
\delta {\cal L}_2= \delta F \wedge v\; i_v(F+*F) - \delta v \wedge v\; (i_v(F+*F))^2 + {1\over 2v^2} F\wedge
\delta F \;  \qquad
 \end{eqnarray}
 and, ignoring the total derivatives, as
\begin{eqnarray}\label{vcL=}
\delta {\cal L}_2= d a\wedge d{\cal G}_0\; (\delta \phi - 2 \delta a \; {\cal G}_0) \; . \qquad
 \end{eqnarray}
Here
 \begin{eqnarray}\label{vcG0=}
 {\cal G}_0= {i_v({ F+*F})\over \sqrt{\partial a
\partial a}}=
{i_v({ d\phi+*d\phi})\over \sqrt{\partial a
\partial a}}
\; . \qquad
 \end{eqnarray}
{}From (\ref{vcL=}) one can clearly read the second  PST symmetry with $\delta \phi = 2 \delta a \; {\cal G}_0$
and arbitrary function $\delta a(t,x)$. But, as far as our basic field is now scalar, the straightforward
counterpart of the first PST symmetry ($\delta B_p=\phi_{p-1}\wedge da$, see the first term in
(\ref{vB=PST+Top}))  is actually absent.

A more careful analysis shows that $\delta \phi= f(a(t,x))$ gives only a total derivative contribution to
(\ref{vcL=}) so that the complete symmetry variation leaving invariant the PST action for the chiral bosons
reads
\begin{eqnarray}\label{vphi=sym}
\delta \phi= 2 \delta a \; {\cal G}_0 +f(a(t,x)) \; . \qquad
 \end{eqnarray}
On one hand, as the function  $f(a(t,x)) $ depends on the 2d spacetime coordinates through its dependence on
the PST scalar only, the second term is the clear counterpart of the semi-local symmetry  of the above Sec.
\ref{sec=semi}. On the other hand this symmetry is present  in the action for chiral bosons  in a topologically
trivial spacetime as well, and, as we will see below, plays the role of PST1 symmetry. This is why  we would
like to call it {\it semi-local PST1 symmetry}. Let us study how it works in the derivation of the chirality
(anti-self-duality) equation (\ref{d+*d=0}) in topologically trivial spacetime $M^{1+1}$.

Eq. (\ref{vcL=}) makes transparent that the  equation of motion which follow from PST action with
 (\ref{cL}) reads
  \begin{eqnarray}\label{dadG0=0}
d a\wedge d{\cal G}_0\equiv d( da\, {\cal G}_0)=0 \; . \qquad
 \end{eqnarray}
In the topologically trivial situation ($b_1=dim \, {\bb H}^1(M^{1+1})=0$) it is solved by
\begin{eqnarray}\label{G0=tf}
{\cal G}_0= \tilde{f}(a(t,x)) \; , \qquad
 \end{eqnarray}
where $\tilde{f}(a(t,x))$ is an arbitrary function of the PST scalar $a(t,x)$.

Now we observe that under the semilocal PST1 symmetry  (\ref{vphi=sym})
\begin{eqnarray}\label{vPST2G0=}
\delta {\cal G}_0= {f}'(a(t,x))\; , \qquad
 \end{eqnarray}
so that {\it r.h.s.} of (\ref{G0=tf}) can be removed or generated by the transformation with  ${f}'(a):={d\over
da}f=\mp \tilde{f}(a)$. Then, {\it if the semilocal PST1 symmetry is a gauge symmetry}, Eq.  (\ref{G0=tf}) is
gauge equivalent to ${\cal G}_0=0$ which in its turn is tantamount to the chirality equation
 (\ref{d+*d=0}),
\begin{eqnarray}\label{G0=0=}
{\cal G}_0= 0\qquad \Leftrightarrow \qquad d\phi+*d\phi=0 \; . \qquad
 \end{eqnarray}
As we will see,  this is the case for {\it da-timelike} branch of the dynamical system described by the PST
action (\ref{cL}) ({\it da-timelike PST}) which is gauge equivalent to the not manifestly Lorentz invariant
Floreanini--Jackiw (FJ) action \cite{Floreanini:1987as},
\begin{eqnarray}\label{SFJ=}
S_{FJ} =- \int dtdx ( \partial_t \phi\partial_x \phi+ \partial_x \phi \partial_x \phi) \; . \qquad
 \end{eqnarray}
 In contrast, for the $da$-spacelike branch of the 2d PST system ({\it da-spacelike PST}), which is gauge
 equivalent to non-manifestly Lorentz invariant 'anti-FJ' or PS-like action,
\begin{eqnarray}\label{SaFJ=}
S_{aFJ} =\int dtdx ( \partial_t \phi\partial_t \phi+\partial_t \phi \partial_x \phi) \; , \qquad
 \end{eqnarray}
 the semi-local PST1 is an infinite dimensional global symmetry.

\subsection{Semi-local symmetry as gauge symmetry of FJ and $da$-timelike PST actions}

Using the PST2 symmetry of the $da$-timelike  PST system to fix the gauge $a(t,x)=t$, where the PST action is
reduced to the FJ action (\ref{SFJ=}), we notice that the 2d counterpart of the PST1 symmetry, the semi-local
PST1 symmetry, is parametrized by a function of time coordinate $x^0=t$ only,
\begin{eqnarray}\label{vphi=0a}
\delta \phi (t,x)= f(t)    \; . \qquad
 \end{eqnarray}
Notice that, if the spacial coordinate takes values in a final interval, $x\in (x_f, x_i)$, (\ref{vphi=0a})  is
also the symmetry of the action provided the scalar field $ \phi (t,x)$ obeys the following boundary conditions
\begin{eqnarray}\label{pxf=pxi} \phi (t,x_f)= \phi (t,x_i)  \; . \qquad
 \end{eqnarray}

The fact that (\ref{vphi=0a}) is a gauge symmetry is intuitively clear. However, it is instructive to prove
this formally.

As it was stressed in recent \cite{Sevrin:2013nca} (see also \cite{Henneaux:1987hz,Henneaux+T=Book})   the
difference between gauge symmetry and infinite dimensional global symmetry, the characteristic example of which
is given by the 2d conformal symmetry,  is that for the former the Noether charges  vanish identically, while
for the latter this is not the case.

Decomposing the parametric function on (\ref{vphi=0a}) in series,  $f(t)=f(0)+tf'(0)+... + {t^n \over n!}
f^{(n)}(0)+...$, and considering $\epsilon^{(n)}=  {f^{(n)}(0)\over n!}$ as the symmetry parameters, we find the
corresponding  Noether currents $J^{\mu (n)}= (J^{0 (n)}, J^{1 (n)})$,
\begin{eqnarray}\label{Jmun=}
&& J^{0 (n)}= -t^n\partial_x\phi \equiv -\partial_x(t^n\phi )  \; , \qquad \nonumber \\ && J^{1 (n)}= -t^n
(\partial_t+2\partial_x)\phi + nt^{n-1} \phi \equiv  \partial_t (t^{n} \phi ) - 2 t^n
(\partial_t+\partial_x)\phi  \; .
 \end{eqnarray}
It is not difficult to see that these currents are conserved on the mass shell, $
\partial_\mu J^{\mu (n)}= \partial_t J^{0 (n)}+ \partial_x J^{1 (n)} = -2t^n \partial_x
(\partial_t+\partial_x)\phi
$, and that the corresponding Noether charges
\begin{eqnarray}\label{QnN=0}
Q^{(n)}= \int dx J^{0(n)}= \int dx \partial_x(t^n\phi )= - t^n (\phi(t, x_f)- \phi (t, x_i))=0
 \;
 \end{eqnarray}
vanish identically with the  boundary conditions (\ref{pxf=pxi}).

Hence, the semilocal PST1 symmetry of the FJ action (\ref{SFJ=}) is gauge symmetry. The same conclusion holds
for the semi-local PST1 symmetry of the $da$-timelike branch of the dynamical system described by the manifestly
Lorentz invariant 2d PST action (\ref{cL}).

\subsection{Chirality equation as gauge fixed form of the Lagrangian equations of the FJ action and of the
da-timelike branch of the PST action}

The Lagrangian equation of motion which follows from the FJ action (\ref{SFJ=}), $\partial_x (\partial_t +
\partial_x )\phi=0$, can be written in the first order form as
\begin{eqnarray}\label{Leq=FJ}
 (\partial_t+\partial_x) \phi=\varphi (t) \; .
 \end{eqnarray}
However, the action (\ref{SFJ=}) and  the second order form of the equations  are invariant under the
semi-local PST1 symmetry (\ref{vphi=0a}) which, as we have shown above, is a gauge symmetry. Choosing $f(t)$ in
(\ref{vphi=0a})  to be a solution of $\dot{f}(t):= \partial_t f(t)=\varphi (t)$ we can gauge away the {\it
r.h.s.} of (\ref{Leq=FJ}) and write this equation in a gauge fixed form
\begin{eqnarray}\label{dt+dx=0}
 (\partial_t+\partial_x) \phi=0 \; .
 \end{eqnarray}
Hence the FJ action (\ref{SFJ=}) can be used to obtain the chirality condition (\ref{dt+dx=0}) as equation of
motion.

This is also the case for the $da$-timelike branch of the PST system. The Lagrangian equations (\ref{dadG0=0})
which follow from the PST action (\ref{cL}) are gauge equivalent to the chirality conditions (\ref{G0=0=}),
$d\phi+*d\phi=0$,  as the semi-local PST1 symmetry is a gauge symmetry in this branch.

\subsection{Issues of anti-FJ action and $da$-spacelike branch of the 2d PST system}

In the case of anti-FJ action (\ref{SaFJ=}) which can be obtained from $da$-spacelike branch of the PST system by
gauge fixing of the PST2 symmetry, the semi-local PST1 symmetry transformation are characterized by a function
of spacial variable $x$,
\begin{eqnarray}\label{vphi=1a}
\delta \phi (t,x)= f(x)    \; .  \qquad
 \end{eqnarray}
A more careful look shows that this symmetry requires the 'initial' conditions
\begin{eqnarray}\label{ptf=pti} \phi (t_i,x)= \phi (t_f,x)  \;  \qquad
 \end{eqnarray}
(in contrast with the boundary conditions (\ref{pxf=pxi}) required for the semi-local symmetry of the FJ
action). The components of Noether currents $J^{\mu (n)}=(J^{0 (n)},J^{1 (n)})$ corresponding to (infinitely
many constant parameters of) this symmetry are
\begin{eqnarray}\label{Jmun=a}
&& J^{0 (n)}= 2 x^n (\partial_t+\partial_x)\phi - \partial_x (x^{n} \phi )\; , \qquad J^{1 (n)}= \partial_t(x^n
\phi)  \; .
 \end{eqnarray}
These currents are conserved on the mass shell and have the  Noether charges
\begin{eqnarray}\label{QnNa=} Q^{(n)}= \int dx J^{0(n)}= - \int dx \partial_x(x^n\phi )=
x_i^n \phi(t, x_i)- x_f^n \phi (t, x_f)\not=0
 \; .
 \end{eqnarray}

Generically, these do not vanish. Hence we conclude that the semi-local PST1 symmetry of the anti-FJ action
(\ref{SaFJ=}) and of the $da$-spacelike branch of the 2d PST system is not a gauge symmetry but an infinite
dimensional rigid symmetry, similar to the famous 2d  conformal symmetry.

That infinite dimensional rigid symmetry cannot be used to gauge away the {\it r.h.s.} of the first order form
of the anti-FJ Lagrangian equation which, thus,  contains an arbitrary function of the spacial coordinate $x$
in {\it r.h.s.}, \begin{eqnarray}\label{Leq=aFJ}
 (\partial_t+\partial_x) \phi=\varphi (x) \; .
 \end{eqnarray}
Of course, this can be written as a chirality equation for the redefined field $\tilde{\phi}=\phi(t,x)-\varphi
(x)$,
 \begin{eqnarray}\label{Leq=taFJ}
 (\partial_t+\partial_x) \tilde{\phi}=0 \; , \qquad \tilde{\phi}=\phi(t,x)-\varphi (x)\; .
 \end{eqnarray}
However, this does not help much as far as the redefinition is done with an arbitrary function and it does not
change the conclusion that the general solution of the equations of motion which follow from the anti-FJ
action,
  \begin{eqnarray}\label{sLeq=aFJ}
 \phi(t,x)= h(t-x) -\varphi (x)\; ,
 \end{eqnarray}
contains, besides  the arbitrary chiral function $ h(t-x)$, also arbitrary function of the spacial variable
$\varphi (x)$.

Thus in the case of 2d chiral bosons, even when 2d spacetime is topologically trivial, the $da$-timelike branch of
the 2d PST system and a non-manifestly Lorentz invariant FJ action, which can be obtained from that by gauge
fixing, are preferable over the $da$-spacelike branch of the 2d PST system and anti-FJ action. Only formers can be
used to obtain the anti-self duality equations,  (\ref{G0=0=}) and (\ref{dt+dx=0}), the general solution of
which are given by one arbitrary function $h(t-x)$. In contrast, for $da$-spacelike PST and anti-FJ action the
general solution of the equations of motion, Eq. (\ref{sLeq=aFJ}), contains also an arbitrary function of the
spacial coordinate, $\varphi (x)$, and in this sense is rather similar to the solution of the equations of
motion of `usual',  non-chiral massless boson.

\subsection{A speculation on alternative canonical formalism in 2d}

In search for possibility to rehabilitate the $da$-spacelike branch of the PST system and the anti-FJ action one
may turn to the idea of \cite{Hull:2014cxa} to develop the canonical formalism using $x$ instead of $t$
variable.

Indeed, the pseudo-Noether-charge assiciated to this formalism,
\begin{eqnarray}\label{tQnNa=0} \tilde{Q}{}^{(n)}= \int dt J^{1(n)}=  \int dt \partial_t(x^n\phi )=
x^n (\phi(t_f, x)- \phi (t_i, x)) =0
 \;
 \end{eqnarray}
vanishes identically as a result of `initial' conditions (\ref{ptf=pti}). As it was noticed in
\cite{Hull:2014cxa}, different choices  of the basic variable of the canonical formalism should correspond to
restrictions to different field configurations, for which the different physically relevant integrals converge. In the
two-dimensional case spacial and temporal slices are one-dimensional so that the exchange of the roles of space
and time variables, and the convergence conditions for spacial and  temporal integrals, does not look
unnatural.

However, after fixing once the basic variable (time) of canonical formalism, one sees that the dynamical system
described by the PST action splits on two branches, $da$-timelike and $da$-spacelike, and that the semi-local PST
symmetry is the gauge symmetry in one, usually chosen to be the first, while it is the infinite dimensional
rigid symmetry in the other. Then
the $da$-timelike branch, the gauge
fixed version of which are given by non-manifestly Lorentz invariant FJ action,  is preferable as it allows to
obtain the chiral boson equation as a gauge fixed version of the Lagrangian equation of motion.

\subsection{Chiral bosons on a Riemann surface }

On Riemann surface $\Sigma_g$ with nonvanishing genus ${\bf g}\not=0$ the first cohomology group ${\bb
H}^{1}(\Sigma_g)$ is $2{\bf g}$ dimensional,  $b_1=2{\bf g} \not=0 $,  so that there exists a basis
$\{\Omega^L_1 \}$ of $2{\bf g}$ closed but not exact forms on  $\Sigma_g$. If $\Sigma_g$ allows for the
existence of a nowhere vanishing vector field, the PST action (\ref{cL}) is well defined   and produces
the Lagrangian equation  (\ref{dadG0=0})\footnote{It is known that if the Riemann surface is {\it compact},
{\it connected}  and admits a nowhere vanishing {\it holomorphic} one-form ((1,0)-form), then it is a torus
(quotient of ${\bb C}$ by a lattice) which implies ${\bf g}=1$.
An exhaustive study of the canonical and BRST quantization of the FJ model on the torus and comparison of the particion function for chiral bosons obtained on this way with the results of  holomorphic factorization approach \cite{Witten:1996hc} can be found in recent \cite{Chen:2013gca}.}.
On $\Sigma_g$  the general solution of
(\ref{dadG0=0})  reads
\begin{eqnarray}\label{daG0=kOm}
da {\cal G}_0=da \, \tilde{f}(a) + k_L\Omega^L_1\; , \qquad d\Omega^L_1=0\; , \qquad \Omega^L_1\not= d \chi
(x)\, , \qquad L=1,..,2{\bf g}\; .
 \end{eqnarray}
It includes $2{\bf g}$ constants $k_L$.

Contracting this equation with $v=da/\sqrt{|\partial a\partial a |}$, we find
 \begin{eqnarray}\label{G0=kivOm}
{\cal G}_0= \tilde{f}(a)\pm  k_L\, i_v\Omega^L_1 /\sqrt{|\partial a\partial a |}\; , \qquad
 \end{eqnarray}
while contracting  the  Hodge dual of Eq. (\ref{daG0=kOm}) gives the following equation for the coefficients
$k_L$:
 \begin{eqnarray}\label{kiv*Om=0}
 k_L\, i_v*\Omega^L_1=0\; . \qquad
 \end{eqnarray}

Similar equations can be derived from non-manifestly diffeomorphism invariant FJ-type action obtained from the
da-timelike branch of PST action by setting $da=dt$. These read
 \begin{eqnarray}\label{G0=ki0Om}
{\cal G}_0:= i_0(d\phi+*d\phi)=
 \tilde{f}(t)+ k_L\, i_0\Omega^L_1 \; , \qquad
 \\ \label{ki0*Om=0}
 k_L\, i_0*\Omega^L_1=0\; \qquad \Leftrightarrow \qquad k_L\, i_1\Omega^L_1=0\; .
 \end{eqnarray}
The last equation restricts the set of constants $k_L$ to be such that $k_L  \Omega^L_1 = dt k_L
i_0\Omega^L_1$. Then the closure of $\Omega^L_1$ forms implies $d(k_L \Omega_1^L)=  dt  \wedge d(k_L
i_0\Omega_1^L)=0$ and hence that $k_L i_0\Omega_1^L$ is $x$--independent, $k_L\, i_0\Omega^L_1=
\tilde{\tilde{f}}(t)$ \footnote{Then the form $k_L  \Omega^L_1 = dt \tilde{\tilde{f}}(t)$ is not exact iff $dt$
is not exact, i.e. iff our time is an angular variable. If not, then the condition that $\Omega^L_1$ forms are
not exact would require to set $\tilde{\tilde{f}}(t)=0$ and we can proceed as in the topologically trivial
case. }. As a result,  Eq. (\ref{G0=ki0Om}) can be written in the form
\begin{eqnarray}\label{G0=ki0Om}
{\cal G}_0:= i_0(d\phi+*d\phi)=
 \tilde{f}(t)+ \tilde{\tilde{f}}(t)\; .
 \end{eqnarray}
Then the second term together with the first one can be gauged away using the semilocal PST1 symmetry
(\ref{vphi=0a}) with $f(t)$ obeying $f'(t)=-\tilde{f}(t)-\tilde{\tilde{f}}(t)$. After this stage we arrive at
the chirality equation (\ref{d+*d=0}).

In the same manner we can reproduce the chirality equation (\ref{d+*d=0}) as a gauge fixed version of the
Lagrangian equations of motion which follow from the PST action (\ref{cL}) written on a Riemann surface
$\Sigma_g$, but only in its {\it da-timelike} branch in which the semi-local PST1 symmetry  in (\ref{vphi=sym})
is a gauge symmetry (see however the speculations in the previous subsection)\footnote{If  $\Sigma_g$ has the
metric of Euclidean signature, then there is no separation of PST system on two branches, all the possible values
of $da$ are related by nonsingular PST2 symmetry transformations, but the chiral boson and the PST Lagrangian
become complex. See sec. 3.5.3 for more discussion in the $D=6$ model. }.

\section{Twisted anti--self--duality of p-form gauge fields from the PST action in a D=2p+2  dimensional
spacetime of nontrivial topology}

The generalization of the  analysis of Sec. 3 to chiral bosons in spacetime of an arbitrary even dimension is
quite straightforward. In this section we will present the basic equations and formulate the conclusions for
the Lagrangian description of chiral p-form gauge fields in topologically nontrivial spacetime of $D=2p+2$
dimensions.

\subsection{Twisted anti-self-duality in $D=2p+2$}

Let us define the measure of $D=2p+2$ dimensional spacetime by $dx^{\nu_1}\wedge \ldots \wedge dx^{\nu_D} =
d^Dx \epsilon^{\nu_1\ldots \nu_D}=-(-)^p d^Dx \epsilon^{\nu_D\ldots \nu_1}$ with $\epsilon^{01\ldots (2p+1)}= -
\epsilon_{01\ldots (2p+1)}=1$, and consider the set of $n$ $p$-form gauge fields $B^I_p$, $ I=1,...,n$,  with
the field strength (see also (\ref{H=dBp+}) below)
\begin{eqnarray}\label{H=dBp}
 H^I_{p+1}= dB_p^I= {1\over (p+1)!} dx^{\nu_{p+1}}\wedge \ldots \wedge dx^{\nu_1}H^I_{\nu_1\ldots \nu_{p+1}}(x)
 \; , \qquad \\ \nonumber  \nu =0,1,...,(2p+1) \; . \qquad
 \end{eqnarray}
If we define the dual of an arbitrary $q$-form $F_{q}= {1\over q!} dx^{\nu_q}\wedge \ldots \wedge
dx^{\nu_1}F_{\nu_1\ldots \nu_q}(x)$  by
\begin{eqnarray}\label{*Fq=}
 *F_q=
 {1\over (D-q)!} dx^{\nu_{D-q}}\wedge \ldots \wedge dx^{\nu_1}\; {\!\sqrt{|g|}\over  q! }\; \epsilon_{\nu_1\ldots
 \nu_{D-1}\mu_1\ldots \mu_{q}} F^{\mu_1\ldots \mu_{q}}(x) \; , \qquad
 \end{eqnarray}
then $**H^I_{p+1}=(-)^pH^I_{p+1}$ so that for odd $p$ (in particular for $p=1$ corresponding to $D=4$) to be
consistent one has to consider a {\it twisted (anti-)self-duality} condition imposed on the even number of
gauge field strengths ($n= 2m$ for $p=2q+1$), rather than just (anti-)self-duality condition. The twisted
anti-self-duality equation has the form
\begin{eqnarray}\label{*cHq=0}
{\cal H}_{p+1}^I= 0\qquad
 \end{eqnarray}
where
\begin{eqnarray}\label{*cHq:=}
{\cal H}_{p+1}^I= H^I_{p+1}+ \Omega_{IJ}*H^J_{p+1} \; . \qquad
 \end{eqnarray}
This contains an invertible  $n\times n$ matrix  $\Omega_{IJ}$ with  the properties
\begin{eqnarray}\label{*cHq:=}
\Omega_{IJ} = (-)^p \Omega_{JI}\; , \qquad \Omega_{IK}\Omega_{KJ}=(-)^p\delta_{IJ}\; , \qquad I,J=1,...,n \; .
 \end{eqnarray}
As a result,
 \begin{eqnarray}\label{cH=Om*cH}
 {\cal H}_{p+1}^I\equiv  \Omega_{IJ}*{\cal H}_{p+1}^J\;
 \end{eqnarray}
 and
 \begin{eqnarray}\label{OmHH=0}
 \Omega_{IJ}H_{p+1}^I\wedge H_{p+1}^J=0\;
 \end{eqnarray}  hold.

\subsection{PST action for chiral bosons  in $M^{2p+2}$}

When spacetime $M^{D}=M^{1+(2p+1)}$ is topologically trivial, it is known that the twisted anti-self duality
equation (\ref{*cHq=0})  can be obtained from the PST action  $S^{PST} \propto \int {\cal L}^{PST}_D$ with
\begin{eqnarray}\label{LPST=D}
{\cal L}_D^{PST}&=& -  \Omega_{IJ} i_v{\cal H}_{p+1}^I \wedge H_{p+1}^J\wedge v = {1\over p!} d^{D} x \, \sqrt{
|g| }
 v^\rho{\cal H}^I_{\rho\mu_1...\mu_p} \; v_\lambda*H^{\lambda\mu_1...\mu_p}\, \;  , \quad \\ \label{v=daD}
& & v=dx^\nu v_\nu = da/\sqrt{|\partial a\partial a|} \; , \qquad \partial a\partial a:= g^{\mu\nu}\partial_\mu
a\partial_\nu a \; . \qquad
\end{eqnarray}
Indeed, modulo exact forms, the variation of ${\cal L}^{PST}_D$ can be written as  \footnote{\label{vcLDfoot}
In terms of the variation of the field strength
\begin{eqnarray}\label{vcLDinfoot}\nonumber
\delta {\cal L}^{PST}_D =  2\Omega_{IJ}  da \wedge {\cal G}_{p}^I \wedge \left( \delta H_{p+1}^J - {1\over 2}
{\cal G}_p^J \wedge d(\delta a) \right) \mp (-)^p \Omega_{IJ}  H_{p+1 }^I\wedge\delta H_{p+1}^J   \; , \qquad
\end{eqnarray}
where the sign of the last term is related to $\pm 1=v_\mu v^\mu$. The property  (\ref{OmHH=0}) is essential to
derive this result. The other useful identities are $\pm F_q=i_vF_q\wedge v+*(i_v*F_q\wedge v)$ and
$i_v*F_q=-(-)^q*(F_q\wedge v)$, in particular, $i_v*H_{p+1}=(-)^p*(H_{p+1}\wedge v)$. }
\begin{eqnarray}\label{vLHH=D}
\delta {\cal L}^{PST}_D =  2(-)^p \Omega_{IJ}  d(da \wedge {\cal G}_{p}^I) \wedge  \left(  \delta B_{p}^J -
\delta a  {\cal G}_p^J \right)   \; , \qquad
\end{eqnarray}
where
\begin{eqnarray}\label{cGI:=D}
{\cal G}_{p}^I := { i_v{\cal H}_{p+1}^I \over \sqrt{|\partial a\partial a|} }= { i_vH^I_{p+1}+
\Omega_{IJ}i_v*H^J_{p+1} \over \sqrt{|\partial a\partial a|}}\; .  \qquad
 \end{eqnarray}
Eq. (\ref{vLHH=D}) makes manifest the PST gauge symmetries
\begin{eqnarray}\label{vBp=PST}
\delta {B}^I_p = \varphi_{p-1}^I  \wedge da + \delta a \; {\cal G}^I_{p}
 \end{eqnarray}
with an arbitrary  $ \delta a (x)$ and $ \varphi^I_{\mu_1 \ldots  \mu_{p-1}} (x)=  \varphi^I_{[\mu_1 \ldots
\mu_{p-1}]} (x)$ in $ \varphi_{p-1}^I ={1\over (p-1)!} dx^{\mu_{p-1}} \wedge \ldots  dx^{\mu_1}
\varphi^I_{\mu_1 \ldots  \mu_{p-1}} (x)$. The presence of the former  shows the pure gauge (St\"{u}ckelberg)
nature of the PST scalar, while the arbitrary $ \varphi^I_{\mu_1 \ldots  \mu_{p-1}} (x)$ allows to gauge away
the general solution $da  \wedge {\cal G}_{p}= da  \wedge d\phi_{p-1}$ of the Lagrangian  equations of motion
\begin{eqnarray}\label{dGpda=0}
d( da  \wedge {\cal G}^I_{p})=0\; , \qquad
\end{eqnarray}
thus arriving at $ {\cal G}_{p}^I=0$ which implies the twisted anti--self--duality equation (\ref{*cHq=0})
\begin{eqnarray}\label{Gp=0}
  {\cal G}^I_{p}=0 \qquad \Rightarrow \qquad  {\cal H}^I_{p}=0\; .
 \end{eqnarray}

Like in $D=6$, for our discussion below   we need to define {\it da--timelike} and {\it da--spacelike} branches,
in which the partial derivative of the PST scalar is spacelike and timelike vector respectively:
\begin{eqnarray}
\label{da-timel}
da-timelike\quad branch\; : \qquad \partial a\partial a:= g^{\mu\nu} \partial_\mu a\partial_\nu a>0\; , \qquad
\\ \label{da-spacel} da-spacelike\quad branch\, :
\qquad
\partial a\partial a:= g^{\mu\nu} \partial_\mu a\partial_\nu a<0\; . \qquad
\end{eqnarray}
In the {\it da--timelike} branch,  the PST scalar  $a(x)$ can be
equated  to the time coordinate by using the smooth local PST2  transformation, while in the {\it
da--spacelike} branch this is impossible but it is possible to equate  $a(x)$ with one of the  space
coordinates, say  $da(x)=dx^1$. Indeed, these two choices,  $da(x)=dx^0$   and  $da(x)=dx^1$, clearly
corresponding to (\ref{da-timel}) and (\ref{da-spacel}),  cannot be related by smooth local PST2
transformations (see sec. 2.2.3 for more details).

\subsection{Twisted anti--self-duality from PST action in topologically nontrivial $M^{2p+2}$}

In the topologically nontrivial spacetime $M^{2p+2}= M^{1+(2p+1)}$ with nonvanishing Bette numbers
 $b_{p}=dim\, {\bb H}^p(M^{2p+2}) \not=0$  and  $b_{p+1}=dim\, {\bb H}^{p+1}(M^{2p+2}) \not=0$  there exist
 $b_p$ closed but not exact $p$-forms
 $\omega^\Lambda_p$ which provide the basis of ${\bb H}^{p}(M^{2p+2})$,
\begin{eqnarray}\label{dompLa=0}
d\omega^\Lambda_p=0\; , \quad \omega^\Lambda_p\not= d\chi_{p-1}^\Lambda\; , \quad \Lambda=1,...,b_p\; ,  \quad
 \end{eqnarray}
 and $b_{p+1}$ closed but not exact $(p+1)$--forms which provide the basis of ${\bb H}^{p+1}(M^{2p+2})$,
   \begin{eqnarray}\label{dOmpL=0}
d\Omega_{p+1}^L=0\; , \qquad \Omega_{p+1}^L(x)\not= d\chi_{p}^L\; , \qquad L=1,...,b_{p+1}\; .
\end{eqnarray}

These latter enter the general solution of the Bianchi identities $dH^I_{p+1}=0$,
\begin{eqnarray}\label{H=dBp+}
 H^I_{p+1}= dB_p^I+ k_L^I\Omega^L_{p+1} \;  \qquad
 \end{eqnarray}
 with constant $k^I_L$'s.

The PST action (\ref{LPST=D}) makes sense in  a topologically nontrivial spacetime  allowing for the existence
of a nowhere vanishing vector field; it can be written for the generalized field strength (\ref{H=dBp+}).
Varying this action  within a fixed topological class, $\delta H_{p+1}=d\delta B_p$, one finds the same equations
(\ref{dGpda=0}) (see sec. 3 for more discussion). However, an equivalent first order representation of these
equations, which can be  obtained as the general solution of (\ref{dGpda=0}) with respect to
${\cal G}^I_{p} \wedge da$, now contains additional topological contributions.

A straightforward generalization of our approach of sec. 3 allows to show the following facts.
\begin{itemize}
\item The first order form of the PST Lagrangian equations  (\ref{dGpda=0}) can be written in the form
\begin{eqnarray}\label{cGpda=G}
{\cal G}^I_{p} \wedge da= -d\phi^I_{p-1}\wedge da  + \check{\omega}^I_{p}\wedge da \; ,  \qquad
\end{eqnarray}
where $\check{\omega}_{p}^I$ are nontrivial solutions of
\begin{eqnarray}\label{dchom=1da}
d\check{\omega}_{p}^I= \check{\omega}^{(1)I}_{p} \wedge da\, ,  \quad d\check{\omega}^{(1)I}_{p}=
\check{\omega}^{(2)I}_{p} \wedge da\, ,  \quad \ldots ,  \quad d\check{\omega}^{(n)I}_{p}=
\check{\omega}^{(n+1)I}_{p} \wedge da\; ,  \quad \ldots \quad
\end{eqnarray}
In the configuration with $da=dt$ the above conditions imply that the  forms $\check{\omega}_{p}^I$ are
`spatially closed' but not `spatially exact', {\it i.e.} obey $d^{(-)}\check{\omega}_{p}^I=0$ and
$\check{\omega}_{p}^I\not=d^{(-)}{\chi}_{p-1}^I$ where $d^{(-)}=d\vec{x}\, \vec{\partial}$ and $\vec{x}$ are
coordinate on the slice $M_t^{2p+1}$ of $M^{2p+2}$.
\item In a spacetime with $b_p\not=0$ a solution of
(\ref{dchom=1da}) is given by $\check{\omega}^I_{p}= l^I_\Lambda (a(x))\omega_p^\Lambda$ were
$\omega_p^\Lambda$ are $b_p$ $p$--forms forming the basis of ${\bb H}^p(M^{2p+2})$, (\ref{dompLa=0}),  and
$ l^I_\Lambda$ are arbitrary functions of  one variables. In the above solution this is taken to be the PST
scalar.
\item At least for particular cases of $M^{2p+2}$ the above
solution is general (a particular example is ${\bb R}\otimes M^{2p+1}$ with $da$ co-tangent to ${\bb R}$) so that
\begin{eqnarray}\label{cGpda=om}
{\cal G}^I_{p} \wedge da= -d\phi^I_{p-1}\wedge da  + l^I_{\Lambda}(a(x)) \omega^\Lambda_p \wedge da(x)\; .
\qquad
\end{eqnarray}
with arbitrary $l^I_{\Lambda}(a(x))$. In this case it becomes
especially transparent that the value of $b_{p}$ (rather than of $b_{p+1}$) is relevant when considering
the PST action for $B_p$ in $M^{2p+2}$.
\item The complete set of symmetries of the PST action is described by (to simplify equations in this item  we omit the superindex $^I$)
\begin{eqnarray}\label{vBp=PST+TopG}
\delta {B}_p =   d\alpha_{p-1}  + \varphi_{p-1}  \wedge da + \delta a \; {\cal G}_{p} +
 \check{\varphi}_p\; , \qquad d\check{\varphi}_p= \check{\varphi}^{(1)}_p \wedge da\; , \qquad
 \end{eqnarray}
with
 \begin{eqnarray}\label{dvp=vp1da}
 d\check{\varphi}_p= \check{\varphi}^{(1)}_p \wedge da \, , \quad  d\check{\varphi}^{(1)}_p=
 \check{\varphi}^{(2)}_p \wedge da \, , \quad \ldots \, , \quad d\check{\varphi}^{(n)}_p=
 \check{\varphi}^{(n+1)}_p \wedge da \; , \quad ... \, . \quad
 \end{eqnarray}
At least in the particular cases (see above) this can be written as
\begin{eqnarray}\label{vBp=PST+Top}
\delta {B}_p = d\alpha_{p-1} + \varphi_{p-1}  \wedge da + \delta a \; {\cal G}_{p} +
 \omega_p^\Lambda \; f_\Lambda(a(x))\; \qquad
 \end{eqnarray}
with $b_p$ arbitrary functions of one variable $f_\Lambda(a)$. These parametrize the semi-local symmetry.
In generic case  the parameters of this semi-local symmetry of the PST action are hidden inside of
$p$--form $\check{\varphi}_p$ which obey the (infinite chain of) equations  (\ref{dvp=vp1da}). \item The
{\it r.h.s.} of the first order form of the Lagrangian PST equations, Eqs.  (\ref{cGpda=G}) or
(\ref{cGpda=om}), can be removed by the standard PST gauge symmetries, described by the second and the
third terms in (\ref{vBp=PST+TopG}) or (\ref{vBp=PST+Top}), and by the semi-local symmetry. \item Thus {\it
if the semi-local symmetry is gauge symmetry}, the first order form of the PST Lagrangian equation,
(\ref{cGpda=G}) or (\ref{cGpda=om}), is gauge equivalent to $ {\cal G}^I_{p}=0$, Eq. (\ref{Gp=0}), which in
its turn is equivalent to the twisted anti-self--duality equation
\begin{eqnarray}\label{*cHq:=H+=0}
{\cal H}_{p+1}^I:=  H^I_{p+1}+ \Omega_{IJ}*H^J_{p+1}=0  \;  \qquad
 \end{eqnarray} for the (field strengths of the) potentials which enter the action.
\item This is the case for the {\it da--timelike} branch of the PST system, and for the HT action which is
    obtained from this by fixing the gauge  $da=dt$. \item For the other,  {\it da}--spacelike branch of the
    PST system, and for the PS action which is obtained from this by fixing the gauge (say) $da=dx^5$, the
    best what one can obtain is the twisted anti-self-duality equation $\tilde{{\cal H}}{}^I_{p+1} :=
    \tilde{{H}}{}^I_{p+1} +  \Omega_{IJ}* \tilde{{H}}{}^I_{p+1}=0 \;$ for redefined field strength,
    $\tilde{{H}}{}^I_{p+1} = {{H}}{}^I_{p+1}  - l^I_{\Lambda}(a(x)) \omega^\Lambda_p \wedge da$
    in the particular cases (see (\ref{cGpda=om}))  and
    $\tilde{{H}}{}^I_{p+1} = {{H}}{}^I_{p+1} - \check{\omega}^I_p \wedge da$ in
    generic spacetime (see (\ref{dchom=1da}) and (\ref{cGpda=G})). As
    far as $da$ can be considered as exact form, this can be interpreted as field strength of redefined
    $p$--form potential,   $\tilde{{B}}{}^I_{p} ={{B}}{}^I_{p} - \tilde{f}^I_{\Lambda}(a(x))
    \omega^\Lambda_p$ in the particular cases and $\tilde{{B}}{}^I_{p} ={{B}}{}^I_{p} -\check{\beta}_p$ with
    $\check{\beta}_p= \check{\omega}_p\wedge da$ in general. Although this redefinition is given by the semi-local
    symmetry transformation, this is not a gauge symmetry of the
      {\it da}--spacelike branch of the PST system, nor of the PS action, so that its parameters should be
      considered as additional degrees of freedom of the dynamical system.
\end{itemize}

To resume, the PST formalism is consistent and can be used to obtain the (twisted anti-)self-duality equations
for $p$-form gauge potentials  also in spacetime $M^{2p+2}$ of nontrivial topology (admitting a nowhere
vanishing vector field). To be more precise, in the spacetime with $b_p\not=0$ and, more generally, in
$M^{2p+2}$ admitting a nontrivial  $p$-forms obeying (\ref{dchom=1da}) this conclusion holds for the  {\it
da-timelike branch} of the dynamical system described by the PST action, as well as for the non-manifestly
invariant HT action which can be obtained from that by gauge fixing.

\section{Conclusions}

\label{Conclusion}

In this paper we have shown that the Pasti--Sorokin--Tonin (PST) approach \cite{Pasti:1996vs,Pasti:1997gx} is
consistent and produces the (twisted anti-)self-duality equation as a gauge fixed version of the equations of
motion also in spacetime of nontrivial topology.

We have began by the basic example of chiral  2-form gauge field in D=6 spacetime $M^{6}$, which has been
elaborated in detail in secs. 2 and 3. This allowed to shorten the presentation of the generic case of chiral
$p$-form gauge fields $B^I_p$ in D=2p+2 dimensional spacetime $M^{2p+2}=M^{1+(2p+1)}$ of nontrivial topology in
Sec. 5. The intermediate Sec. 4 is devoted to the special case of $D=2$ chiral bosons.

The PST action contains an auxiliary scalar field $a(x)$ which is pure gauge ( St\"{u}ckelberg  field) with
respect to a specific  gauge symmetry (PST2 gauge symmetry). However, as far as ${1\over \sqrt{|\partial
a(x)\partial a(x)|}}$ enters the action and the Lagrangian equations, not all the configurations of  $a(x)$ are
allowed. We stress that this topological restriction implies the existence of two branches of the dynamical
system described by the PST action (PST system): {\it da-timelike} branch in which the PST scalar can be gauged to
coincide with time coordinate, or better to say $da=dt$,   and {\it da-spacelike} branch in which the gauge
$da=dx^1$ is accessible.

In the gauge $da=dt$ the ({\it da-timelike} branch of the) PST action reduces to the (non-manifestly Lorentz
invariant) Henneaux--Teitelboim (HT) action \cite{Henneaux:1988gg,Henneaux+T=Book}. In D=2 such an action was
discussed in \cite{Floreanini:1987as} by Floreanini and Jackiw so that we call this FJ action.  In the gauge
$da=dx^1$ the ({\it da-spacelike} branch of the) PST action reduces to another non-manifestly Lorentz invariant
functional which in D=6 was considered by Perry and Schwarz  \cite{Perry:1996mk}; we call this PS action while
for its D=2 counterpart we also use the name anti-FJ action.

The PST action can be written in any curved spacetime $M^{2p+2}=M^{1+(2p+1)}$  provided it allows for the
existence of a nowhere vanishing vector field. However, in some case the topology intervenes  the process of
derivation of (twisted anti)-self duality equations from the Lagrangian equations of the PST action. Namely,
this happens if  $M^{2p+2}=M^{1+(2p+1)}$  allows for the existence of nontrivial $p$--forms $\check{\omega}_p$
which obey $d\check{\omega}_p=\check{\omega}^{(1)}_p\wedge da(x)$, where $da(x)$ is an arbitrary nowhere
vanishing closed 1--form (which could be exact and identified with the derivative of the PST scalar field) and
$\check{\omega}^{(1)}_p$ is implicitly defined by the same equation. If $M^{2p+2}$ allows for the existence of
$b_p$ linearly independent closed but not exact $p$--forms $\omega_p^\Lambda$
 ($\Lambda =1,..., b_p$) at least a particular class of such $\check{\omega}_p$ is provided by
 $\sum\limits_{\Lambda=1}^{b_p}f_\Lambda (a(x)) \omega_p^\Lambda$, where $f_\Lambda (a)$ are arbitrary
 functions of one variable.

In this case the first order form of the PST Lagrangian equation acquires an additional contribution to its
{\it r.h.s.} and, on the first glance, are not gauge equivalent to the (twisted anti--)self duality equation.
However, a more careful study shows that in such spacetimes the PST action also possesses an additional {\it
semi-local symmetry} and that the additional terms in the {\it r.h.s.} of the first order form of the
Lagrangian PST equations can be removed or generated by the transformations of this semi-local symmetry.
Furthermore, we have shown that {\it for the da-timelike branch} of the PST system this semi-local symmetry is a
gauge symmetry and, hence the additional terms in right hand side can be gauged away reducing the Lagrangian
equations to the (twisted-anti-)self duality equations.

In the other {\it da--spacelike} branch of the PST system the semi-local symmetry is an infinite dimensional
rigid symmetry, similar to the conformal symmetry in 2d, which cannot be used to remove degrees of freedom. As
a result, although the Lagrangian PST equations in this branch can be written as   (twisted-anti-)self duality
equations for a redefined potential, and the redefinition can be identified with semi-local symmetry
transformations, the parameters of these should be considered as parameters of the general solution of the
equations of motion and thus as additional degrees of freedom making the content of the model different from
just chiral boson(s).

As we have commented in the main text, it is tempting, following \cite{Hull:2014cxa}, to speculate on a
hypothetical possibility to improve the situation with {\it da--spacelike} branch of the PST system by
developing an alternative canonical formalism which uses one of the spacial coordinate instead of time.
However, if one would like to deal with two branches of the PST system simultaneously, one should
use the same formalism for both, so that our problem remains for one of two branches. This provided us with an additional reason to keep in this paper a more conservative point of view and to stay within the standard canonical formalism.

Thus, curiously enough, the topology makes difference between $da$-timelike and $da$-spacelike branches of the PST
system making the first preferable as its equations of motion are gauge equivalent to the (twisted-anti-)self
duality equations and,  hence, the field content in this branch is given by one (or several) chiral boson(s).

An important problem is to understand the implications of our results for quantum theory of D-dimensional
chiral bosons.

Another interesting issue is  the influence of the spacetime topology on the generalized PST approach of
\cite{Pasti:2009xc,Ko:2013dka} with several PST scalars: $a^r=(a^1,...,a^q)$ with $q>1$ ($q=3$ in
\cite{Pasti:2009xc,Ko:2013dka}).  Instead of ${1\over \sqrt{|\partial a(x)\partial a(x)|}}$ the generalized PST
action and equations of motion would include the inverse $Y_{rs}^{-1}$ of the matrix $Y^{rs}= g^{\mu\nu}(x)
\partial_\mu a(x)^r \partial_\nu a^s(x)$. Hence the requirement for spacetime manifold to have a nowhere
vanishing vector field will be replaced in this case by the requirement of a nowhere singular $q\times q$
matrix $Y^{rs}= g^{\mu\nu}(x) \partial_\mu a(x)^r \partial_\nu a(x)$, $det Y^{rs}\not= 0$, or,  equivalently,
of the nowhere singular rank q projector $P_{\mu}{}^\nu = \partial_\mu a(x)^r  Y_{rs}^{-1}\partial^\nu a^s(x)$.
(This can be formulated as requirement of  the existence of nowhere singular  q-plane field). We hope to address this problem in near future.

{\bf Acknowledgements}. The author is grateful to Dima Sorokin for collaboration at early stages of this
project, numerous useful discussions and comments on final version of the manuscript.  He thanks Marc Henneaux for useful conversation. This work was supported in part by research grant
FPA2012-35043-C02-01 from  MINECO of Spain, by the Basque Government research group grant ITT559-10, and by
UPV/EHU under the program UFI 11/55.

\renewcommand{\thesection}{A}

\renewcommand{\theequation}{A.\arabic{equation}}
\section{On first order form of the Lagrangian equations of the PST system in flat spacetime}

Here we present some details on the derivation of the first order form (\ref{cG2da=dp1da}) of the Lagrangian
equations (\ref{dG2da=0}) which follow from the PST action (\ref{LHH=}) in topologically trivial 6D spacetime.

At first glance, it seems that one can solve (\ref{dG2da=0}) by a more general expression
\begin{eqnarray}\label{cG2da=p2da}
{\cal G}_2\wedge da = \phi_2 \wedge da\; ,\qquad d\phi_2= \phi^{(1)}_2\wedge da\; , \qquad
\end{eqnarray}
where  $d\phi^{(1)}_2= \phi^{(2)}_2\wedge da$ etc. However, we will  see that in the case of topologically
trivial spacetime this does not go beyond the solution (\ref{cG2da=dp1da}).

For simplicity, let us discuss the case $da=dt$, when the PST action reduces to the HT action. Let us define
the splitting $\phi_2= \phi_2^{(-)} + i_0\phi_2 \wedge dt$, $d= d^{(-)}+dt\partial_t$ {\it etc.}. Then   Eq.
(\ref{cG2da=p2da}) can be written in the form
\begin{eqnarray}\label{cG2dt=p2dt}
{\cal G}_2\wedge dt = \phi_2^{(-)} \wedge dt\; ,\qquad
\end{eqnarray}
where $d\phi_2^{(-)}= \phi^{(1)(-)}_2\wedge dt$. This last equation can be equivalently written as
$d^{(-)}\phi_2^{(-)}= 0$. As far as we are in topologically trivial spacetime, its spacial part is also
topologically trivial so that $d^{(-)}\phi_2^{(-)}= 0$ is solved by $\phi_2^{(-)}=d^{(-)} \phi_1^{(-)}$. This
can be equivalently written as  $\phi_2^{(-)}=d \phi_1-  (\partial_t \phi_1^{(-)}- d^{(-)} i_0\phi_1) \wedge
dt$.  Clearly, only the first term contributes to the {\it r.h.s.} of  (\ref{cG2dt=p2dt}) which, hence, can be
equivalently written in the form of Eq. (\ref{cG2da=dp1da}), 
\begin{eqnarray}\label{AcG2da=dp1da}
{\cal G}_2\wedge dt = d\phi_1 \wedge dt= - d(\phi_1 \wedge dt) \; . \qquad
\end{eqnarray}

\renewcommand{\thesection}{B}
\renewcommand{\theequation}{B.\arabic{equation}}
\section{Noether currents and Noether charges in a 6d theory of 2-form gauge potential}

The variation of the 6d action $\int {\cal L}_6$ for the 2-form potential $B_2$ can be written as
\begin{eqnarray}\label{vL6=gen}
&\int_{M^6}\delta {\cal L}_6= \int_{M^6} \left( {\delta {\cal L}_6\over \delta B_2}\wedge \delta B_2+ {\delta
{\cal L}_6\over \delta H_3}\wedge d\delta B_2\right) = \int_{M^6}  {\cal E}_4 \wedge \delta B_2+ \int_{M^6}
d\left( {\delta {\cal L}_6\over \delta H_3}\wedge \delta B_2\right)\; ,  \qquad \nonumber \\
\end{eqnarray}
where
\begin{eqnarray} \label{vL6Eq=gen}
{\cal E}_4= {\delta {\cal L}_6\over \delta B_2}-  d{\delta {\cal L}_6\over \delta H_3}\;  \qquad
\end{eqnarray}
is the {\it l.h.s.} of the Lagrangian equations of motion. A transformations of the 2-form potential
\begin{eqnarray} \label{vB2=aym-gen}
\delta_\varphi B_2=  R_{2 {\cal A}} \varphi^{{\cal A}} \;  \qquad
\end{eqnarray}
with constant parameters $\varphi^{{\cal A}}$ and field dependent 2-forms $R_{2 {\cal A}} =R_{2 {\cal A}}(B_2)$
is a symmetry if $\delta_\varphi {\cal L}_6= \varphi^{{\cal A}} dK_{5 {\cal A}}$. As, on the other hand, Eq.
(\ref{vL6=gen}) implies $\delta_\varphi {\cal L}_6= {\cal E}_4 \wedge R_{2 {\cal A}} \varphi^{{\cal A}} +
d\left({\delta {\cal L}_6\over \delta H_3}\wedge R_{2 {\cal A}} \right)$, we see that the 5-form
\begin{eqnarray} \label{*J1=gen}
*J_{1 {\cal A}}= {\delta {\cal L}_6\over \delta H_3}\wedge R_{2 {\cal A}} - K_{5 {\cal A}} \qquad
\end{eqnarray}
is closed on the mass shell,
\begin{eqnarray} \label{d*J1=0}
d*J_{1 {\cal A}}= 0 \qquad when \qquad {\cal E}_4=0\; . \qquad
\end{eqnarray}
The Noether current $J^\mu_{{\cal A}}$ can be defined by
\begin{eqnarray} \label{J1=gen}
*J_{1 {\cal A}}=: {1\over 5!} dx^{\mu_5} \wedge \ldots \wedge dx^{\mu_1} \epsilon_{\mu_1\ldots \mu_5 \mu}
J^\mu_{{\cal A}}\; \qquad
\end{eqnarray}
and Eq. (\ref{*J1=gen}) is equivalent to its  conservation,
\begin{eqnarray} \label{d*J1=0}
\partial_\mu J^\mu_{{\cal A}}= 0 \qquad when \qquad {\cal E}_4=0\; .
\qquad
\end{eqnarray}

For a gauge symmetry the 5-form ($(D-1)$--form) dual to the Noether current is not only closed, but exact on
the mass shell so that the corresponding Noether charge vanishes identically.

As an example let us consider the standard 2-form gauge symmetry $\delta B_2= d\alpha_1$ of the action with
Lagrangian form dependent on $B_2$ through $H_3=dB_2$ only. In this case the equations of motion read
$d\left({\delta {\cal L}_6\over \delta H_3}\right)=0$. Rewriting the gauge  transformations in the form of
$\delta B_2= dx^\mu \wedge dx^\nu \; \sum\limits_{n=0}^{\infty}{1\over n!}x^{\rho_n} \ldots
x^{\rho_1}\partial_{\rho_1}\ldots \partial_{\rho_1}\partial_{[\nu}\alpha_{\mu]}$ and identifying the constant
parameters as $\{ \varphi^{{\cal A}}\}=\{ {1\over n!}\partial_{\rho_1}\ldots
\partial_{\rho_1}\partial_{[\nu}\alpha_{\mu]}\}$,  we find the 5-forms dual to  Noether currents
\begin{eqnarray} \label{*Jsg=}
*J^{\nu_1\nu_2\; \rho_1 \ldots \rho_n}_{1}= {\delta {\cal L}_6\over \delta H_3}\wedge \left( dx^{\nu_1} \wedge
dx^{\nu_2} \; x^{\rho_1} \ldots x^{\rho_n}\right)_{\begin{matrix} {}^{\Box\Box\ldots\ldots\Box}_{\Box}
\end{matrix}}  \; . \qquad
\end{eqnarray}
The graphical subscript ${\begin{matrix} {}^{\Box\Box\ldots\ldots\Box}_{\Box}
  \end{matrix}}$  in this expression indicates that one should take only one irreducible part of the tensorial
  2-form in the brackets, that corresponding to the Young diagram  represented by the subscript \footnote{The
  tensorial 2-form
 $dx^\mu \wedge dx^\nu \; x^{\rho_n} \ldots x^{\rho_1}$ carries reducible representation of the Lorentz group
$\begin{matrix} {}^{\Box}_{\Box} \end{matrix} \otimes   \begin{matrix} {}^{\Box\ldots\ldots\Box}
  \end{matrix} = {\begin{matrix} {}^{\Box\Box\ldots\ldots\Box}_{\Box}
  \end{matrix}} \oplus {\begin{matrix} {}^{\Box}_{\Box} & {}^{\ldots\ldots\Box} \cr {}^{\Box} &
  \end{matrix}}$. }. This extraction of one irreducible part allows to conclude that
   $\left( dx^{\nu_1} \wedge dx^{\nu_2} \; x^{\rho_1} \ldots x^{\rho_n}\right)_{\begin{matrix}
   {}^{\Box\Box\ldots\ldots\Box}_{\Box}   \end{matrix}}=\propto
  d \left( dx^{[\nu_1} x^{\nu_2]} \; x^{\rho_1} \ldots x^{\rho_n}\right)_{\begin{matrix}
  {}^{\Box\Box\ldots\ldots\Box}_{\Box}   \end{matrix}}$ and, thus, that, on the mass shell,   the 5-form dual
  to Noether current is exact
\begin{eqnarray} \label{*Jsg=}
*J^{\nu_1\nu_2\; \rho_1 \ldots \rho_n}_{1}=\propto d\left( {\delta {\cal L}_6\over \delta H_3}\wedge \left(
dx^{[\nu_1}x^{\nu_2]} \; x^{\rho_1} \ldots x^{\rho_n}\right)_{\begin{matrix}
{}^{\Box\Box\ldots\ldots\Box}_{\Box}   \end{matrix}} \right) \; . \qquad
\end{eqnarray}
This is tantamount to saying that the Noether current is given by the divergence of an antisymmetric  tensor,
$J^{\mu \; \nu_1\nu_2\; \rho_1 \ldots \rho_n}=
 \partial_{\mu^\prime }(...)^{\mu^\prime\mu \; \nu_1\nu_2\; \rho_1 \ldots \rho_n}_{1}$, so that the Noether
 charge vanishes,
 $Q^{\mu \; \nu_1\nu_2\; \rho_1 \ldots \rho_n}= \int d^5 x
 J^{0 \; \nu_1\nu_2\; \rho_1 \ldots \rho_n}= \int d^5 x
 \partial_{i}(...)^{i0 \; \nu_1\nu_2\; \rho_1 \ldots \rho_n}_{1} =0$ (we do not consider here the possible
 boundary contributions).

As a second example, we can consider the PST1 gauge symmetry, or better its counterpart for the
Henneaux--Teitelboim action $\delta B_2=\phi_1\wedge dt= -dt\wedge dx^i\phi_i(t,\vec{x})$, for which the
constant parameters are $-{1\over n! }\partial_{\mu_1}\ldots \partial_{\mu_n}\phi_i(0,\vec{0})$ and the formal
expression for the 5-forms dual to the Noether currents read
\begin{eqnarray} \label{*Jsg=}
*J^{i\; \mu_1 \ldots \mu_n}_{1}= {\delta {\cal L}_6\over \delta H_3}\wedge  dt \wedge dx^{i} \; x^{\mu_1}
\ldots x^{\mu_n}  \; . \qquad
\end{eqnarray}
However, for the Henneaux--Teitelboim action which have the above described symmetry, ${\delta {\cal L}_6\over
\delta H_3}={\cal G}_2\wedge dt$ so that the Noether current vanishes identically, $*J^{i\; \mu_1 \ldots
\mu_n}_{1}\equiv 0$.

The same conclusion follows for the  $\delta B_2=\phi_1\wedge dx^5$ of the Perry--Schwarz action, confirming
that this is also a gauge symmetry characterized by vanishing Noether current.

\end{document}